\documentclass[]{IEEEtran}
\usepackage{amsmath}
\usepackage{amsthm}
\usepackage{amsfonts}
\usepackage{amssymb}
\usepackage{makeidx}
\usepackage{cite}
\usepackage{graphicx,bigints}
\usepackage{mathtools}
\usepackage{cases}
\usepackage{color}
\usepackage{hyperref}
\usepackage{bbm}
\usepackage[normalem]{ulem}
\usepackage{braket}
\usepackage{epstopdf}
\usepackage{xcolor}
\usepackage{lipsum}
\usepackage{kotex}

\usepackage{multirow}

\newtheorem{theorem}{Theorem}
\newtheorem{corollary}{Corollary}
\newtheorem{lemma}{Lemma}

\newtheorem{example}{Example}

\newtheorem{remark}{Remark}

\DeclareMathOperator*{\argmin}{arg\,min}

\DeclareMathOperator{\cA}{\mathcal{A}}

\DeclareMathOperator{\cT}{\mathcal{T}}

\DeclareMathOperator{\cL}{\mathcal{L}}

\DeclareMathOperator{\SIR}{\textrm{SIR}}

\DeclareMathOperator{\bP}{\mathbf{P}}

\DeclareMathOperator{\bE}{\mathbf{E}}
\DeclareMathOperator{\bZ}{\mathbb{Z}}
\newcommand*\diff{\mathop{}\!\mathrm{d}}

\newcommand*\nnb{\nonumber}

\definecolor{sandy}{HTML}{E6E2AF}
\definecolor{stone}{HTML}{A7A37E}
\definecolor{beach}{HTML}{EFECCA}
\definecolor{ocean}{HTML}{046380}
\definecolor{diver}{HTML}{002F2F}

\definecolor{Firenze1}{HTML}{468966}
\definecolor{Firenze2}{HTML}{FFF0A5}
\definecolor{Firenze3}{HTML}{FFB03B}
\definecolor{Firenze4}{HTML}{B64926}
\definecolor{Firenze5}{HTML}{8E2800}
\definecolor{mediumpersianblue}{rgb}{0.0, 0.4, 0.65}
\definecolor{hongik}{HTML}{004498}
\definecolor{cobalt}{rgb}{0.0, 0.28, 0.67}
\definecolor{burntorange}{rgb}{0.8, 0.33, 0.0}

\definecolor{ultramarineblue}{rgb}{0.25, 0.4, 0.96}

\title{Analysis of a Spatially Correlated Vehicular Network Assisted by Cox-distributed Vehicle Relays}

\author{Chang-Sik Choi and François Baccelli
	\IEEEcompsocitemizethanks{\IEEEcompsocthanksitem{Chang-Sik Choi is with Hongik University, South Korea. François Baccelli is with Inria Paris and with Telecom Paris, France.}(email: chang-sik.choi@hongik.ac.kr, francois.baccelli@inria.fr). This paper is an extension of our early work \cite{9870681}.}
}
\begin{document}
	\maketitle 
	\begin{abstract}
	In vehicle-to-all (V2X) communications, roadside units (RSUs) play an essential role in connecting various network devices. In some cases, users may not be well-served by RSUs due to congestion, attenuation, or interference. In these cases, vehicular relays associated with RSUs can be used to serve those users. This paper uses stochastic geometry to model and analyze a spatially correlated heterogeneous vehicular network where both RSUs and vehicular relays serve network users such as pedestrians or other vehicles. We present an analytical model where the spatial correlation between roads, RSUs, relays, and users is systematically modeled via Cox point processes. Assuming users are associated with either RSUs or relays, we derive the association probability and the coverage probability of the typical user. Then, we derive the user throughput by considering interactions of links unique to the proposed network. This paper gives practical insights into designing spatially correlated vehicular networks assisted by vehicle relays. For instance, we express the network performance such as the user association, SIR coverage probability, and the network throughput as the functions of network key geometric variables. In practice, this helps one to optimize the network so as to achieve ultra reliability or maximum user throughput of a spatially correlated vehicular networks by varying key aspects such as the relay density or the bandwidth for relays. 
	\end{abstract}

	\section{Introduction}
	\subsection{Background and Motivation}
	
	Recent innovations have made it possible for vehicles to play new roles in urban environments, extending their traditional transportation role \cite{5888501,6702523,6823640}. Vehicles will participate in various road safety and efficiency applications by communicating with neighboring vehicles, pedestrians, traffic lights, and Internet-of-Things (IoT) devices \cite{5888501,6823640}. Advanced vehicles and their sensors provide ways to improve not only their own safety, but also that of others such as pedestrians \cite{36885,7992934}. This innovative use of vehicles requires reliable communications among network entities such as vehicles, base stations, smart sensors, and pedestrians \cite{7992934,22816,22836}. 
	
	\par Vehicular networks featuring reliable and high capacity links can be achieved by base stations close to roads, namely RSUs. Connected to the core network through backhaul, RSUs will host advanced V2X applications \cite{36885,9345798}. As the number of network users increases and vehicular networks have more services, some network users {relying only on RSUs} may experience limited coverage because of data congestion, load unbalancing, signal attenuation, and high interference. 
	\par
\begin{figure}
		\centering
		\includegraphics[width=1\linewidth]{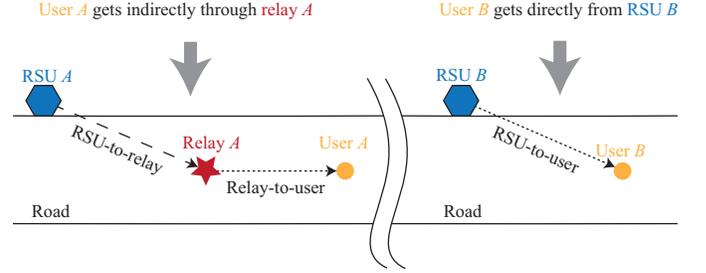}
		\caption{Illustration of the proposed vehicular network with RSUs, relays, and users. Users may get messages directly from RSUs (right) or via relays (left). }
		\label{fig:basicheterovehicle}
	\end{figure}

 To fight against these limitations, various technologies have been developed and among that, this paper focuses on the use of vehicular relays \cite{38836,38874}. Specifically, RSU-operated vehicular relays will reshape the topology of vehicular networks to increase the reliability and throughput of network \cite{38836,38874}. For instance, users in dense areas can communicate each other via relays, avoiding extra delays occurring at RSUs; or relays can forward important messages to the users far away from RSUs. Fig. \ref{fig:basicheterovehicle} illustrates such an example.  

Focusing on the network topology and the geometric interaction between network elements, this paper studies the fundamental performance of a spatially correlated two-tiered heterogeneous vehicular network with RSUs and relays operated through them. (Fig. \ref{fig:basicheterovehicle}.) In particular, to describe the spatial interaction between RSUs, relays, and users all at the same time, we employ a stochastic geometry framework \cite{baccelli1997stochastic,daley2007introduction,chiu2013stochastic,5226957}. In particular, many studies used analytic models based on the Poisson point processes \cite{1580787,6042301,6171996,6497002} where network elements can be well captured as spatially independent components. Recently,  Cox point processes has been employed to describe the locations of spatially correlated network elements such as roads, vehicles, and pedestrians.  Specifically, the Cox models were extensively used in various papers including \cite{morlot2012population,8340239,8357962,8419219,8353411, 8796442,Choi2018Densification,9201475,9354063,9477118,9524528,9782581,9792580,10024366,10066317} to analyze the basic network performance with vehicle transmitters and vehicle receivers. 

Continuing this line of work, this paper also employs Cox point processes to describe the locations of RSUs, users, and vehicle relays, all within the same road infrastructure. It is worth noting that, due to this practical representation of spatially correlated network elements---RSUs, vehicular relays, and users, the network performance improved by vehicle relays can be fairly and accurately analyzed. To the best of the authors' knowledge, no prior work has addressed a system-level analysis of a vehicular network with vehicle relays, especially by emphasizing the network topology produced by RSUs, vehicle relays, and network vehicle users, all of which must be located on the common road infrastructure.

	\subsection{Theoretical Contributions}
\subsubsection{Modeling of a spatially correlated two-tier heterogeneous vehicular network} In vehicular networks, network elements e.g., RSUs, vehicles transceivers, and pedestrians are all close to roads. In this paper, we focus on this geometric characteristic by modeling a road layout first as a Poisson line process and then distributing RSUs, vehicular relays, and users as Poisson point processes conditional on the line process. By constructing all elements conditionally on roads, we account for the fact that RSUs, relays, and users are located on roads and nowhere else. The proposed network modeling technique allows one to identify the geometric interaction of a two-tier heterogeneous vehicular network, especially between various communication links such as RSU-to-relay links, relay-to-user links, and RSU-to-user links. In contrast to our previous work \cite{8419219,8357962} where only a single set of vehicle transmitters is considered as a Cox point process, this work considers two sets of transmitters modeled as Cox point processes conditionally on a single road layout.

\subsubsection{Association behavior of users and coverage probability} Motivated by basic safety messages transmitted from network elements and received by nearby users \cite{37885,38211}, we assume that network users are associated with their closest RSUs or closest relays. We derive the association probability as a function of relay density and RSU density. The obtained probability describes the fraction of users associated with RSUs or with relays, at any given time. We show that the association probability is not given by a simple linear function because of the spatial correlation between RSUs and RSU-operated relays. Assuming frequency resources are separated for operating relays and for serving network users, we evaluate the coverage probability of the typical user as an integral function. See \ref{S:II} for detail. 

\subsubsection{Comprehensive analysis and design insights} Taking into account the fact that relays are operated by RSUs and users are served by both relays and RSUs, we evaluate the effective throughput of the typical user in the proposed network. In particular, we get the user throughput formula leveraging (i) the throughputs of RSU-to-user links and relay-to-user links, respectively, (ii) the SIR distribution and throughput of RSU-to-relay links, (iii) the average number of network elements involved in the above links. Without ignoring the bottleneck resulting from the RSU-to-relay links, the throughput formula accurately describes the redistribution of the network payload achieved by spatially correlated relays in heterogeneous vehicular network architectures. In particular, we express the user throughput as a function of network parameters including frequency resources and densities of RSUs, relays, and users. As a result, it can be effectively used to design and build heterogeneous vehicular networks where spatially correlated network elements exist. For instance, leveraging the throughput expression, network operators can allocate frequency resources to various links to optimize the network performance for given densities of RSUs, relays, and network users.

%	\textbf{Heterogeneous vehicular architecture}: At a high level, this paper studies a multi-tier heterogeneous vehicular network leveraging a Cox point process \cite{baccelli1997stochastic,morlot2012population,8419219}. A multi-tier cellular network stochastic geometry model was proposed and analyzed in \cite{6171996} where heterogeneous base stations are modeled as independent planar Poisson point processes. In contrast to \cite{6171996}, our work studies heterogeneous vehicular network where the transmitters---RSUs and vehicular relays---are modeled as Cox point processes conditional on the same line process. To the best of the authors' knowledge, the present paper is a first attempt to model and analyze the spatially correlated network elements present in heterogeneous vehicular networks. Table \ref{tablefirst}  depicts the stochastic geometry models for the class of heterogeneous networks considered in the present paper.  

%
%\begin{table}
%	\centering
%	\caption{Stochastic geometry model for heterogeneous networks}\label{tablefirst}
%	\begin{tabular}{|c|c|}
%			\hline
%			Network  & Spatial model \\\hline
%		 	Cellular \cite{6171996} & Poisson point processes \\\hline
%			Vehicular \cite{8357962} & Poisson and Cox point processes \\\hline
%			Vehicular [This paper] & Cox and Cox point processes\\\hline
%		\end{tabular}
%\end{table}

	\section{System Model}\label{S:II}
	This section gives the spatial model for RSUs, relays, and users. We then discuss the propagation model, the user association principle, and performance metrics.

	\subsection{Spatial Model for RSUs and Users}
	
	To represent road geometries in urban areas, we assume that the road layout is modeled as an isotropic Poisson line process $  \Phi  $ \cite{chiu2013stochastic}. In the context of stochastic geometry such a model has been widely accepted for its analytical tractability \cite{8340239,8357962,8353411, 8796442,Choi2018Densification,9201475,9354063,9477118}. Specifically, the Poisson line process is generated from a homogeneous Poisson point process on a cylinder set $ \mathbf{C} $. Consider a Poisson point process $ \Xi  $ of intensity $ \lambda_l/\pi $ on $ \mathbf{C} $. Here $ \lambda_l $ (per km) is the mean number of road segments in a circle of radius $ 1 $ km. 
	
	Each of its point $ (r,\theta) $ is mapped into a line on the Euclidean plane, where $ r $ corresponds to the distance from the origin to the line and $ \theta $ corresponds to the angle between the line and the $ x $-axis, measured in the counterclockwise direction \cite{chiu2013stochastic}. 
	
%	
%	\begin{figure}
%		\centering
%		\includegraphics[width=0.6\linewidth]{roadlay_v3}
%		\includegraphics[width=0.6\linewidth]{roadlay_v2}
%		\caption{Illustration of RSUs, relays, and users in the proposed network.  The road layout is the Poisson line process. RSUs, relays, and users are the Cox point processes conditional on the same line process. Here, $ \lambda_l=3 $/km. $ \mu_s = 2 $/km, $ \mu_r= 5 $/km, $ \mu_u=10 $/km. Colored polygons represent building surfaces and walls in urban areas. }
%		\label{fig:deployment2}
%	\end{figure}
%	
	
	Conditionally on each line $ l(r,\theta) \in  \Phi , $ the locations of RSUs and network users, are modeled as independent one-dimensional Poisson point processes $ S_{r,\theta} $ and $U_{r,\theta} $ of intensities $ \mu_s $ and $ \mu_u $, respectively, where $ \mu_u\gg\mu_s  $. Here, $ \mu_s $ is the mean number of RSUs on a road segment of $ 1 $ km and $ \mu_r $ is the mean number of relays on a road segment of $ 1 $ km. 
	
	Collectively, the RSU point process $ S $ and the user point process $ U $ form Cox point processes constructed under an identical Poisson line process \cite{8419219}. We have 
	\begin{align}
		S   &= \bigcup\limits_{r_i,\theta_i\in \Phi }S_{r_i,\theta_i},\label{1}\\
	U  &= \bigcup\limits_{r_{i},\theta_{i}\in \Phi } U_{r_i,\theta_i}. \label{2}
	\end{align}

\begin{figure}
	\centering
	\includegraphics[width=1\linewidth]{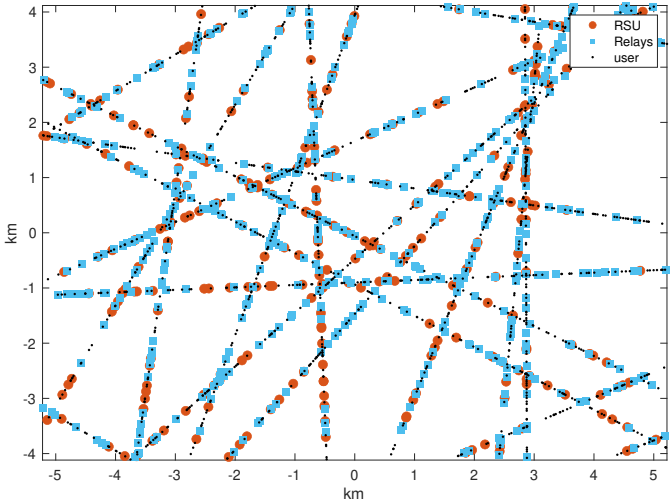}
	\caption{Illustration of the proposed network where $ \lambda_l=2/\text{km}$, $ \mu_s=2/\text{km}$, $ \mu_r= 4/\text{km}$, and $ \mu_u=10/\text{km} $. }
	\label{fig:newdeployment3}
\end{figure}

\begin{figure}
	\centering
	\includegraphics[width=1\linewidth]{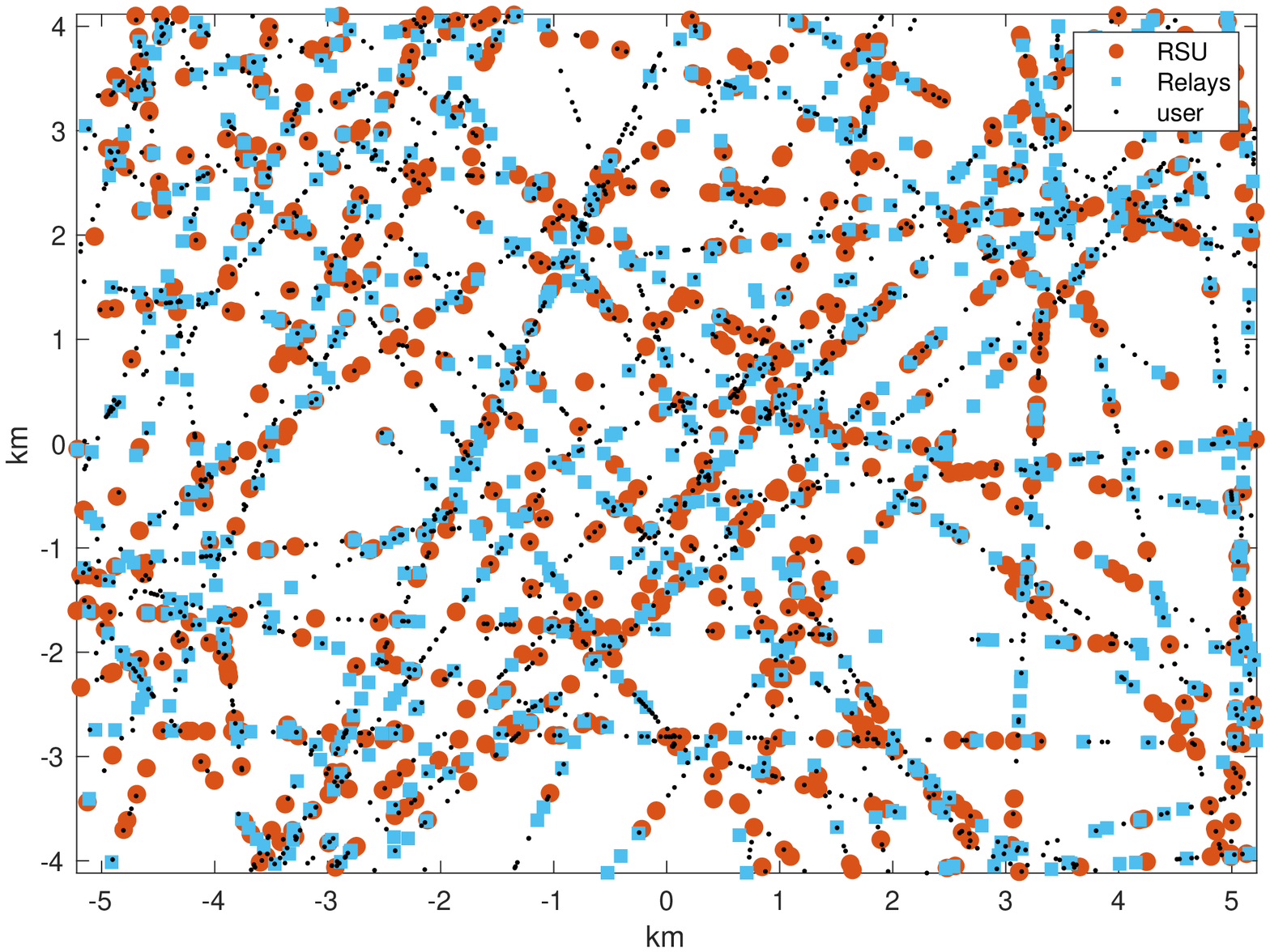}
	\caption{Illustration of the proposed network where $ \lambda_l=5/\text{km}$, $ \mu_s=2/\text{km}$, $ \mu_r= 2/\text{km}$, and $ \mu_u=5/\text{km} $. }
	\label{fig:newdeployment2}
\end{figure}

\begin{figure}
	\centering
	\includegraphics[width=1\linewidth]{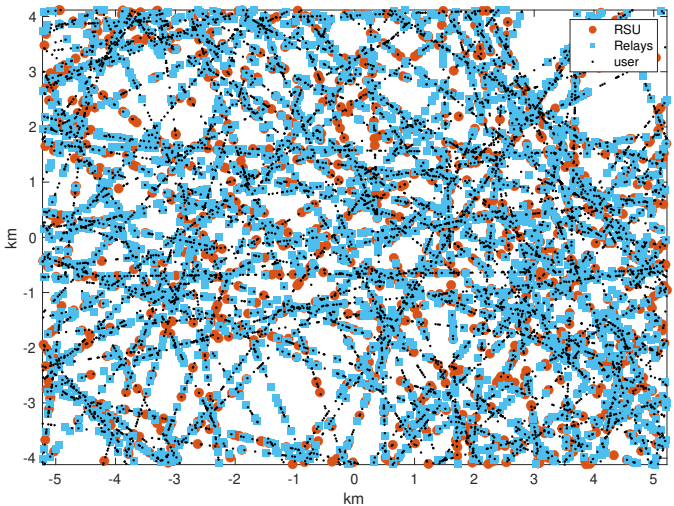}
	\caption{Illustration of the proposed network where $ \lambda_l=10/\text{km}$, $ \mu_s=2/\text{km}$, $ \mu_r= 4/\text{km}$, and $ \mu_u=10/\text{km} $. }
	\label{fig:newdeployment1}
\end{figure}

	Figs. \ref{fig:newdeployment3} -- \ref{fig:newdeployment1} show the proposed network model having RSUs, vehicle relays, and network users, all located on the same road infrastructure. In Fig. \ref{fig:newdeployment3}, the road density is $2$/km and it means there are $4$ roads in a disk of radius $1$ km on average. In Fig. \ref{fig:newdeployment1}, we use $\lambda_l=10$, a very dense urban area with many roads. It is worth noting that RSUs, relays, and network users are all constrained by the common road infrastructure.

	\begin{remark}
		It is important to mention that we take the the simplest approach to characterize spatially correlated component in a two-tier heterogeneous vehicular network. For instance, $ \Xi $ is a homogeneous Poisson point process of a constant intensity. Thus, we have an isotropic Poisson line process $  \Phi  $ on the Euclidean plane. Nevertheless, one can change it by considering dirac-delta measure across the angle of roads to create a Manhattan-like road layout. See \cite{8357962}. A more realistic model obtained by changing the intensity measure of $ \Xi $ is left for future work.
	\end{remark}

%
%	\begin{table}
%	\centering
%	\caption{Network elements}\label{table:1}
%	\begin{tabular}{|c|c|}
%		\hline
%		Variable	& Description\\
%		\hline
%		$  \Phi  $	& Road layout: 2-D Poisson line \\
%		\hline
%		%			$ S_{r,\theta} $	& RSU Poisson point process on $ l(r,\theta) $  of intensity $ \mu_s $ \\\hline 
%		%			$ R_{r,\theta} $	& Relay Poisson point process on $ l(r,\theta) $  of intensity $ \mu_r $\\\hline
%		%			$ U_{r,\theta} $	& User Poisson point process on $ l(r,\theta) $ of intensity $ \mu_u $ \\\hline
%		$  S   $	& RSU Cox point process \\
%		\hline
%		$ R $ 	& Relay Cox point process \\
%		\hline
%		$  U $	& User Cox point process \\
%		\hline
%		$ \lambda_l $ & Road density  per km \\ \hline
%		$ \mu_s $ & RSU density per km \\\hline
%		$ \mu_r $ & Relay density per km \\\hline
%		$ \mu_u $ & User density per km \\\hline
%	\end{tabular}
%	
%\end{table}

	\subsection{Spatial Model for Relay and Reserved Spectrum}\label{S:24}
	\par We assume that vehicular relays are wirelessly connected to RSUs and they serve network users \cite{38211,38836} as in Fig. \ref{fig:basicheterovehicle}. In the sequel, RSU-operated vehicular relays will be referred to relays. 
	
	Since relays are on roads too, we model the locations of relays as a Cox point process, denoted by $ R $. Specifically, conditional on each road $ l(r,\theta) $ created by the above Poisson line process $ \Phi $, the locations of relays on each road follow a Poisson point process $ R_{r,\theta} $ of intensity $ \mu_r. $ Following the notation of Eqs. \eqref{1} and \eqref{2}, we let 
		\begin{equation}
		  R  = \bigcup\limits_{r_i,\theta_i\in \Phi } R_{r_i,\theta_i}.
	\end{equation}
It is important to note that the RSU point process $ S $, the relay point process $ R $, and the user point process $ U $ are all on the same line process $ \Phi $. As a result, our approach captures the fact that RSUs, relays, and users are all on the very same road structure.  Fig. \ref{fig:newdeployment3} shows the spatial distributions of the RSUs, relays, and network users in the proposed network.

 To operate relays, network operators can employ various approaches. To maintain the tractability of our work, we consider the simplest assumption that the frequency resources for \emph{operating} relays and the frequency resources for \emph{serving} network users are separate. (See Fig. \ref{fig:basicheterovehicle} where links are shown.) This is partly motivated by the radio resource management technique in practice  \cite{38211,38836}, where the frequency resources can be autonomously taken by vehicles or scheduled by RSUs. Specifically, to communicate with relays, RSUs use the spectrum $ f_2 $ of bandwidth $ W_2 $. On the other hand, to serve network users on roads, RSUs and relays use the spectrum $ f_1 $ of bandwidth $ W_1$. In other words, we have three different types of communication links (i) RSU-to-user links, (ii) relay-to-user links, and (iii) RSU-to-relay links. Types (i) and (ii) use $ W_1$ and Type (iii) uses $W_2$. 
 
 Note that we assume that $ f_1 $ and $ f_2 $ do not overlap and that $ W_1+W_2=W $ where $ W $ is the total available bandwidth. Table \ref{table:2} shows the communication links and their corresponding resources. 

 	\begin{remark}
 	In practical cases, users may experience limited coverage because of interference or attenuation. In the proposed architecture, RSUs configure relays to forward their messages to network users. To ensure reliable reception of messages at their final destinations, we assume that the initial links of such relaying, namely RSU-to-relay communications occupy a reserved spectrum $ f_2 $ of bandwidth $ W_2 $. For the rest of the communication links in the proposed two-tier heterogeneous vehicular networks, e.g., relay-to-user and RSU-to-user links, we consider those links use a spectrum $ f_1 $ of bandwidth $ W_1 $. Therefore, there is no co-channel interference between RSU-to-relay communications and the rest of the communications in the proposed architecture. Motivated by current standard implementation \cite{38211,38213,38836}, we assume that RSU-to-user and relay-to-user links exist on the same spectrum and thus there is co-channel interference between them.
  \end{remark}	
 	\begin{table}
	\centering
	\caption{Spectrum usage}\label{table:2}
	\begin{tabular}{|c|c|}
		\hline
		Communication link types	& Bandwidth \\
		\hline
		RSU-to-user links & $ W_1 $ \\\hline
		relay-to-user links & $ W_1 $ \\\hline
		RSU-to-relay links & $W_2$ \\\hline
	\end{tabular}
	
\end{table}

	\subsection{Relay and User Mobility}
In vehicular networks, RSUs do not move while relays and network users move along the roads. We assume that relays and users move along the line they are located on and that they choose their speeds uniformly out of a given distribution. Specifically,  each relay independently selects its own speed on the interval $ [v_{\text{r;min}},v_{\text{r;max}}] $ uniformly at random. Each network user selects its own speed on the interval $ [v_{\text{u;min}},v_{\text{u;max}}] $ uniformly at random.

	\begin{example}
One can relax the above mobility assumption. An example is that where relays and users on each road choose their own speeds out of standard normal distributions. Based on the displacement theorem \cite{baccelli2010stochastic}, the Poisson property of the relay and user point processes is preserved over time. Thus, the relay and user point process are still given by time invariant Cox point processes. This shows that the proposed mobility model and the corresponding analysis in this paper generalize to various mobility cases.
	\end{example}

\subsection{Relay Association and User Association}\label{S:relay}
	With regards to relays, we assume that relays are associated with their closest RSUs. The RSU-to-relay communication links are established between RSUs and relays; and then, these relays forward messages from RSUs to nearby users. See Fig. \ref{fig:basicheterovehicle}.  Combined with the separate spectrum usage given in \ref{S:24}, the nearest association is a basis for the reliable reception of forwarded messages at the final destinations.
	
With regards to users, we assume that each user is associated with its closest transmitter, namely either an RSU or a relay. This is based on practical use cases \cite{22816,22836,36885,37885} where network users are configured to connect with their nearest transmitters. The bottom figures of Fig. \ref{fig:newdeployment3} shows the user association map as the Voronoi tessellation, illustrated by solid blue lines. The centers of the Voronoi cells are the transmitter point process, i.e., $ S+R $. The cells are the user association map. Users are connected to transmitters at their cell centers. 

As the number of transmitters increase, the average size of cells decreases and thus the average number of users associated with each transmitters. 
	
	\begin{remark}
		\cite{9870681} studied various user association techniques including the maximum average receive signal power association and the nearest user association. In this paper, motivated by the various distance-critical safety V2X applications, we focus on the nearest user association principle. Nevertheless, the formulas and analysis given in this paper can be readily used to analyze the maximum average receive signal power association simply by changing the coefficients of transmit powers, exploiting techniques in \cite{6171996,6287527,6678062}.
	\end{remark}

	\subsection{Propagation Model} \label{S:prop}
	Consider a receiver located at a distance $ d $ from its transmitter. In the proposed heterogeneous vehicular network, transmitters are either RSUs or relays. We assume rich scattering around the network users \cite{goldsmith2005wireless} and a power-law path loss function \cite{37885}. The received signal power at the receiver is assumed to be of the form 
$		pHL(d)$
	where $ p =\{ p_s , p_r \}$ is the received signal powers at $1$ meter from an RSU or a relay, respectively.  $ H $ represents Rayleigh fading, modeled by an independent exponential random variable with average one, and $ L(d)$ is the path loss over distance $ d\geq 1 $. We assume that the transmit powers for RSUs and relays are given by $  p_s  $ and $  p_r  $, respectively. 
	
	For path loss, we address that the path loss shows different characteristics, depending on the relative locations of transmitters and receivers, or more precisely, on whether transceivers are on the same road or not \cite{38901}. For tractability, we use a simple path loss model where the path loss $ L(d)   $ over a distance $ d $ is 
	\begin{align}L(d)=
		\begin{cases}
			d^{-\alpha} & \text{on the same road},\\
			d^{-\beta}  & \text{on different road},
		\end{cases}\label{pathloss}
	\end{align} 
	where $ 2 < \alpha \leq \beta$. % Thanks to the above assumption, we identify and analyze cross-road and inter-road interference present in the proposed network.  
	
%	{The communications involving users, namely RSU-to-user links and relay-to-user links use wireless frequency resources of $ W_1 $. We assume that RSU-to-relay links use a disjoint wireless spectrum of $ W_2 $. Hence, RSU-to-relay communications do not interfere with RSU-to-user or relay-to-user communications. }

%TODO relay association picture is removed for page constraints. We can add it later 
%\begin{figure}
%	\centering
%	\includegraphics[width=.5\linewidth]{Experiment/roadlay_v4}
%	\caption{Illustration of the relay association region. Here, $ \lambda_l=3 $/km. $ \mu_s = 2 $/km, and $ \mu_r= 5 $/km.  Relays are associated with the closest RSUs. The relay association regions---dotted lines---correspond to the Voronoi tessellation with respect to the RSU point process. Note that it is not the same as the user association region characterized by the Voronoi tessellation with respect to the RSU plus relay point process. }
%	\label{fig:roadlayv4}
%\end{figure}

	\subsection{Performance Metrics}\label{S:perf}
	This paper analyzes the performance seen by the network users. We first derive the coverage probability of the typical user and then derive the coverage probability of the typical relay. Then, using both, we derive the user effective throughput.

	\subsubsection{User Coverage Probability}

	To analyze the coverage probability of the typical user, we use the Palm distribution of the user point process, $ \bP_{ U }^0 (\cdot)$. This features a typical user at the origin. Therefore a line $ l(0,\theta_0) $ almost surely exists with a RSU point process $ S_{0,\theta_0}, $ and a relay point process $ R_{0,\theta_0} $ on it  \cite{morlot2012population,8419219}. The coverage probability of the typical user $ \bP_{ U }^0(\SIR>\tau) $ is given by 
	\begin{align}
		\bP_{ U }^0\left(\frac{pHL(\|X^\star\|)}{\sum_{X_j\in S+R \setminus B(\|X^\star\|)} p_{X_j} H_j L ( \|X_j \|)   }>\tau \right)\label{SIR},
	\end{align}
	where $ p $ is the transmit power of the association transmitter which could be either $ p_s $ or $ p_r $ depending on the user association. We denote by $ B_0(\|X^\star\|) $ the  ball of radius $ \|X^\star\| $ centered at the origin; we also write $ B_0(r)\equiv B(r). $ Here, $ \tau $ is the SIR threshold. Based on the association principle of above, the association transmitter $ X^\star $ is given by 
	 \begin{equation}
		X^\star = \argmin_{X_k\in S_{0,\theta}+R_{0,\theta}+ S + R } \|X_k\|.
	\end{equation} Here, the association transmitter is selected out of the point processes: $ S_{0,\theta_0}, $ $ R_{0,\theta_0}, $ $  S  $, and $  R . $  When the association transmitter is a RSU, we write $ X^\star=X_S^\star $. When the association transmitter is a relay, we write $ X^\star= X_R^\star. $

	\par Based on the association, users can be divided into two types, namely those associated with RSUs and those with relays. We need to separately evaluate each type as follows: \begin{align}
		\bP_U^0(\SIR_{S\to U}>\tau):=&\bP_{ U }^0(\SIR>\tau| X^\star = X_S^\star)\label{SIR_relay},\\
		\bP_U^0(\SIR_{R\to U}>\tau):=&\bP_{ U }^0(\SIR>\tau|X^\star = X_R^\star)\label{SIR_RSU},
	\end{align}
	where the former denotes the coverage probability of the typical relay-associated user and the latter denotes that of the typical RSU-associated user.

	\subsubsection{Relay Coverage Probability} 
	To analyze the SIR of the typical relay, we consider the Palm distribution of the relay point process. The coverage probability of the typical relay, $ \bP_{ R }^0(\SIR_{S\to R}>\tau) $ is given by 
	\begin{align}				
		\bP_{ R }^0\left(\frac{p_sHL(\|X_S^\star\|)}{\sum_{X_j\in S  \setminus B(\|X_S^\star\|)} p_s H_j L ( \|X_j \|)  }>\tau \right)\label{SIRRtor},
	\end{align}
	where $ X_S^\star $ is the RSU closest to the typical relay located at the origin under the Palm distribution of the relay point process $ R $. Since RSU-to-relay communications are assumed to occur over a frequency bandwidth of $ W_2 $, it is worth noting that the RSU-to-relay links do not interfere with RSU-to-user and relay-to-user links. 
	\subsubsection{Throughput} Using the coverage probabilities above, we derive the throughput of the typical user. In the proposed vehicular network where relays are operated by RSUs over a separate wireless resource $ W_2, $ the throughput is not just a simple function of the SIR distribution of the typical user. The precise definition of the user throughput will be given in Section \ref{S:34}.  
\section{Association Probability}\label{S:3}

%This section shows the association probability of the typical user. 

% 

Each user has either a RSU association or a relay association, depending on its distances to RSUs and relays. Here, we study the probability that the typical user is associated with either an RSU or a relay. The association probability is derived under the Palm distribution of the user point process. The association probability also corresponds to the fraction of network users associated with RSUs and with relays, respectively.

\begin{figure*}
		\begin{align}
		\bP(A_s)=&\int_{0}^{\infty}2\mu_s\exp\left(-2(\mu_s+\mu_r) r-2\lambda_l \int_0^r 1-e^{-2(\mu_s+\mu_r)\sqrt{r^2-u^2}}\diff u\right)\diff r\nnb\\ &+4\mu_s\lambda_l\int_{0}^\infty\int_{0}^{\pi/2} r e^{-2r(\mu_s+\mu_r) -2r(\mu_s+\mu_r) \sin(\theta)-2\lambda_l\int_0^r 1-\exp{\left(- 2(\mu_s+\mu_r) \sqrt{r^2-u^2}\right)}\diff u}\diff \theta \diff r.\label{156}
	\end{align}
\rule{\linewidth}{.2mm}
	
\end{figure*}
\begin{lemma}\label{Theorem:2}
	The probability that the typical user is associated with an RSU is given by Eq. \eqref{156} 
	Likewise, the probability that the typical user is associated with a relay is $ \bP( A_r  ) = 1- \bP( A_s).  $ 
\end{lemma}
\begin{IEEEproof}
	See \cite[Theorem 1]{9870681}.
\end{IEEEproof}

\begin{figure}
	\centering
	\includegraphics[width=1\linewidth]{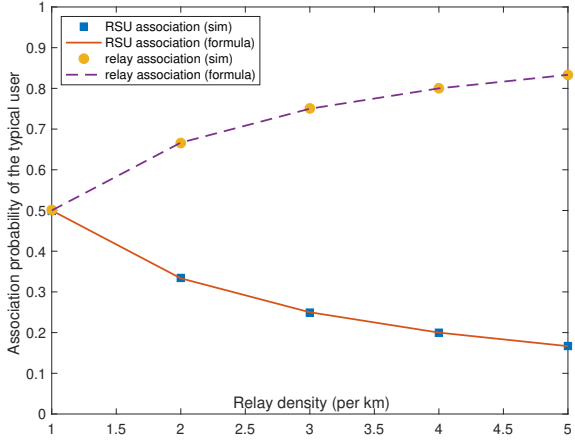}
	\caption{Illustration of the association probability of the typical user. The derived formula of Lemma \ref{Theorem:2} matches the simulation results. We use $ \lambda_l=2 $/km and $ \mu_s=1 $/km.}
	\label{fig:association1115}
\end{figure}

Fig. \ref{fig:association1115} shows that the derived association probability derived in Lemma \ref{Theorem:2} matches the association probability numerically obtained by the Monte Carlo simulations. We use $ \lambda_l=2/km $. We see that as the density of relays increases, the relay association probability increases. Since the user association is based on distance, it is possible that users are associated with RSUs or relays on different lines. We will use the derived association probability to evaluate the coverage probability of the typical user.  In addition, we show that the association probability is not given by a simple ratio of densities. This contrasts to the association probability of users in the heterogeneous networks modeled by Poisson point processes \cite{6171996}. This occurs because of the spatial correlation between RSUs and relays. 
%\begin{example}
%	If we assume the payloads of users are the same, the total amount of payload processed by RSUs is proportional to the number of users associated with RSUs. For example, when $ \mu_s=1 $ and $ \mu_r=0, $ all users are associated with the RSUs. Theorem \ref{Theorem:2} shows that the number of users associated with RSUs can be reduced by a factor of two when we employ as many relays as RSUs. To reduce the load of the RSUs by a factor of four, one needs to deploy a number of relays about three or four times greater than the number of RSUs. Such a redistribution of the payload from RSUs to relays gives a diminishing return as the number of relays increases. In practice,  a network operator could use this formula to estimate the number of relays needed to achieve a certain level of payload redistribution in such heterogeneous vehicular networks.
%\end{example}

\begin{theorem}\label{prop:1}
	The mean number of users associated to the typical RSU is $ \frac{\mu_u}{\mu_s} \bP_u(A_s)$ and the mean number of users associated to the typical relay is $ \frac{\mu_u}{\mu_r}\bP_u^0(A_r). $ The mean number of relays associated to the typical RSU is $ \frac{\mu_r}{\mu_s}. $
\end{theorem}
\begin{IEEEproof}
	Consider a factor graph with an edge from each user to its association RSU or relay. From the mass transport principle \cite{baccelli2010stochastic},  
		\begin{equation}
			\lambda_l\mu_u \bP_U^0(A_s,E) = \lambda_l\mu_s d_{\text{in}}\label{M1},
		\end{equation}
	where the left-hand side is the mean mass sent by the users to their association RSUs on the same lines, whereas the right-hand side is the mean mass received by the RSUs from their associated users on the same lines. $ \lambda_l\mu_s $ is the spatial density of RSUs and $ d_{\text{in}} $ is the mean number of same-line users associated to the typical RSU under the Palm distribution of $ S $. Similarly, considering users and their associated RSUs on different lines, we have   
	\begin{equation}
		\lambda_l\mu_u \bP_U^0(A_s,E^c) = \lambda_l\mu_s d_{\text{in}}',\label{M2}
	\end{equation}where the left-hand side is the mean mass out of the users and the right-hand side is the mean mass received by the RSUs:  $ d_{\text{in}}' $ is the mean number of different-line users associated to the typical RSU.  As a result, the mean number of users per RSU is $d_{in}'+d_{in}={\mu_u}\bP_U^0(A_s) /{\mu_s}$. Similarly, the mean number of users per relay is $ {\mu_u}\bP_U^0(A_r)/{\mu_r}. $ Finally, the mean number of relays per RSU is ${\mu_r}/{\mu_s} $. 
\end{IEEEproof}
The above proposition is essential to address the impact of RSU-to-relay links to the system performance. We use the above expression in the derivation of the user throughput in Section \ref{S:34}.

\section{Coverage Probability of User and Relay}\label{S:31} 
In Section \ref{S:32}, we first evaluate the coverage probability of the typical user under the Palm distribution of the user point process, by leveraging the facts that the network users are connected with their closest RSUs or relays and that RSU-to-user links interfere with relay-to-user links and vice versa. Then, in Section \ref{S:33} we independently derive the coverage probability of the typical relay under the Palm distribution of the relay point process. The coverage probabilities of Sections \ref{S:32} and \ref{S:33} impacts the throughput of the network that we will see in Section \ref{S:34}. 
\subsection{Coverage Probability of the Typical User}\label{S:32} 
This section gives the coverage probability of the typical user. Note all RSUs and relays are assumed to have users to serve with high probability. We denote by $ \gamma $ the ratio of relay transmit power to the RSU transmit power, $ \gamma = p_r/p_s. $ As in Section \ref{S:3}, let $ E $ be the event that the association transmitter and the typical user are on the same line. We denote by $ A_s $ the event that the association transmitter is an RSU and by $ A_r $ the event that the association transmitter is a relay.

\begin{figure*}
	\begin{align}
		&\left.\int_0^\infty G_1(r,a,b)e^{-2\lambda_l\int_0^r1-G_2(r,v,a,b) \diff v-2\lambda_l\int_r^\infty 1-G_3(r,v,a,b) \diff v}\diff r\right|_{a=1,b=\frac{1}{\gamma}} .\label{eq:1}\\
		&\left.\int_0^\infty G_1(r,a,b)e^{-2\lambda_l\int_0^r1-G_2(r,v,a,b) \diff v-2\lambda_l\int_r^\infty 1-G_3(r,v,a,b) \diff v}\diff r\right|_{a=\gamma,b={1}} .\label{eq:2}\\
		&\int_{0}^\infty H_1(r,a,b)e^{-2\lambda_l\int_0^r 1-H_2(r,v,a,b)\diff v}\left.e^{-2\lambda_l\int_r^{\infty} 1-H_3(r,v,a,b)\diff v}H_4(r,c)\diff r\right|_{a=1,b=\frac{1}{\gamma},c=\mu_s} .\label{eq:3}\\
		&\int_{0}^\infty H_1(r,a,b)e^{-2\lambda_l\int_0^r 1-H_2(r,v,a,b)\diff v}
		\left.e^{-2\lambda_l\int_r^{\infty} 1-H_3(r,v,a,b)\diff v}H_4(r,c)\diff r\right|_{a=\gamma,b={1},c=\mu_r}.\label{eq:4}
	\end{align}
	\rule{\linewidth}{0.2mm}
\end{figure*}
\begin{theorem}\label{Theorem:1}
	 The coverage probability of the typical user is 
$ 		\bP_{U}^0(\SIR>\tau,E,A_s)+\bP_{U}^0(\SIR>\tau,E,A_r)
		+\bP_{U}^0(\SIR>\tau,E^c,A_s)+\bP_{U}^0(\SIR>\tau,E^c,A_r), $ given by Eq. \eqref{eq:1} -- \eqref{eq:4}, respectively where 
		% The first term is the coverage probability with the association RSU being on the line $ l(0,\theta_0) $, and the second term is the coverage probability with the association relay being on the line $ l(0,\theta_0) $. Similarly, the third term is the coverage probability with the association RSU not being on $ l(0,\theta_0) $ and the last term is the coverage probability with the association relay not being on $ l(0,\theta_0) $. For Eq. \eqref{eq:1} -- \eqref{eq:2}, we have 
	\begin{align*} 
	&G_1(r,a,b)=2\mu_s e^{-2r\mu_s-2\mu_s\int_{r}^\infty \frac{\tau r^\alpha u^{-\alpha}}{a+\tau r^\alpha u^{-\alpha}}\diff u} \nnb\\
	&\hspace{20mm}e^{-2\mu_r\int_{r}^\infty \frac{\tau r^\alpha u^{-\alpha}}{b+\tau r^\alpha u^{-\alpha}}\diff u},
	\end{align*}
	\begin{align*}
	&G_2(r,v,a,b)=e^{-2(\mu_s+\mu_r)\sqrt{r^2-v^2}} \nnb\\
	&\hspace{25mm}e^{-2\mu_s \int_{\sqrt{r^2-v^2}}^{\infty}\frac{\tau r^\alpha {(v^2+u^2)}^{-\frac{\beta}{2}}}{a+\tau r^\alpha {(v^2+u^2)}^{-\frac{\beta}{2}}}\diff u}\nnb\\
	&\hspace{25mm}e^{-2\mu_r \int_{\sqrt{r^2-v^2}}^{\infty}\frac{\tau r^\alpha {(v^2+u^2)}^{-\frac{\beta}{2}}}{b+\tau r^\alpha {(v^2+u^2)}^{-\frac{\beta}{2}}}\diff u},
\end{align*}
\begin{align*}
		&G_3(r,v,a,b)=e^{-2\mu_s \int_{0}^{\infty}\frac{\tau r^\alpha {(v^2+u^2)}^{-{\beta}/{2}}}{a+\tau r^\alpha {(v^2+u^2)}^{-{\beta}/{2}}}\diff u}\nnb\\
	&\hspace{24mm}e^{-2\mu_r \int_{0}^{\infty}\frac{ \tau r^\alpha {(v^2+u^2)}^{-{\beta}/{2}}}{b+\tau r^\alpha {(v^2+u^2)}^{-{\beta}/{2}}}\diff u}.
\end{align*}
On the other hand, we also have 
	\begin{align*} 
	&H_1(r,a,b)=\! e^{{-2\mu_s r}- 2\mu_s \int_r^\infty \frac{\tau r^\beta u^{-\alpha}}{a+\tau r^\beta u^{-\alpha}}\diff u}\nnb\\
	&\hspace{20mm}e^{- 2\mu_r \int_r^\infty \frac{\tau r^\beta u^{-\alpha}}{b+\tau r^\beta u^{-\alpha}}\diff u},
	\end{align*}
		\begin{align*} 
	&H_2(r,v,a,b) = e^{-2(\mu_s+\mu_r)\sqrt{r^2-v^2}}\nnb\\
	&\hspace{20mm}e^{-2\mu_s \int_{\sqrt{r^2-v^2}}^{\infty}\frac{\tau r^\beta {(v^2+u^2)}^{-{\beta}/{2}}}{a+\tau r^\beta {(v^2+u^2)}^{-{\beta}/{2}}}\diff u} \nnb\\
	&\hspace{20mm}e^{-2\mu_r \int_{\sqrt{r^2-v^2}}^{\infty}\frac{ \tau r^\beta {(v^2+u^2)}^{-{\beta}/{2}}}{b+\tau r^\beta {(v^2+u^2)}^{-{\beta}/{2}}}\diff u},
		\end{align*}
	\begin{align*} 
	&H_3(r,v,a,b)=  e^{-2\mu_s \int_{0}^{\infty}\frac{\tau r^\beta{(v^2+u^2)}^{-{\beta}/{2}}}{a+\tau r^\beta {(v^2+u^2)}^{-{\beta}/{2}}}\diff u}\nnb\\
	&\hspace{20mm}e^{-2\mu_r \int_{0}^{\infty}\frac{ \tau r^\beta{(v^2+u^2)}^{-{\beta}/{2}}}{b+\tau r^\beta {(v^2+u^2)}^{-{\beta}/{2}}}\diff u},
		\end{align*}
	\begin{align*} 
		&H_4(r,c)= \int_{0}^{\pi/2}4\lambda_l c r e^{-2r(\mu_s+\mu_r) \sin(\theta)}\nnb\\
		&\hspace{20mm}e^{-2c\int_{r\sin(\theta)}^\infty \frac{\tau r^\beta(r^2\cos^2(\theta)+v^2)^{-\beta/2}}{1+\tau r^\beta(r^2\cos^2(\theta)+v^2)^{-\beta/2}}\diff v}\diff \theta.
\end{align*}
\end{theorem}

 \begin{figure*}
 		\begin{align}
 		&\prod_{r_i,\theta_i}^{\Phi +\delta_{r_\star,\theta_\star}+\delta_{0,\theta_0}}\left(\bE_{}\left[\prod_{T_j\in S_{0,0}}^{|T_j|>\sqrt{r^2-r_i^2}}\frac{1}{1+p_s s L(\|r_i \vec{e}_{i,1} +T_j \vec{e}_{i,2}\|)}\right]\bE_{}\left[\prod_{T_j\in R_{0,0}}^{|T_j|>\sqrt{r^2-r_i^2}}\frac{1}{1+p_r s L(\|r_i \vec{e}_{i,1} +T_j \vec{e}_{i,2}\|)}\right]\right).\label{eq:mei}\\
		 &4\lambda_l\mu_sr\int_{0}^{\pi/2}e^{-2r(\mu_s+\mu_r) \sin(\theta)-\int_{r\sin(\theta)}^\infty \frac{2\mu_s s  p_s  (r^2\cos^2(\theta)+v^2)^{-\frac{\beta}{2}}}{1+s  p_s  (r^2\cos^2(\theta)+v^2)^{-\frac{\beta}{2}}} -\frac{2\mu_rs  p_r  (r^2\cos^2(\theta)+v^2)^{-\frac{\beta}{2}}}{1+s  p_r  (r^2\cos^2(\theta)+v^2)^{-\frac{\beta}{2}}}\diff v}\diff \theta.\label{eq:233}\\
		&4\lambda_l\mu_rr\int_{0}^{\pi/2}e^{-2r(\mu_s+\mu_r) \sin(\theta)-\int_{r\sin(\theta)}^\infty \frac{2\mu_s s  p_s  (r^2\cos^2(\theta)+v^2)^{-\frac{\beta}{2}}}{1+s  p_s  (r^2\cos^2(\theta)+v^2)^{-\frac{\beta}{2}}} -\frac{2\mu_rs  p_r  (r^2\cos^2(\theta)+v^2)^{-\frac{\beta}{2}}}{1+s  p_r  (r^2\cos^2(\theta)+v^2)^{-\frac{\beta}{2}}}\diff v}\diff \theta.\label{eq:23}\\
				&\int_0^\infty J_1(r)e^{-2\lambda_l\int_{0}^r 1-J_2(r,v)\diff v}e^{-2\lambda_l\int_r^\infty 1-J_3(r,v)\diff  v}\diff r + \int_{0}^\infty K_1(r)e^{-2\lambda_l\int_0^r 1-K_2(r,v)\diff v-2\lambda_l\int_r^{\infty} 1-K_3(r,v)\diff v}K_4(r)\diff r.\label{eq:39}
\end{align}
	\rule{\linewidth}{.1mm}
\end{figure*}

In above, we derived the coverage probability of the typical user at the origin. Yet, the result is applicable to all the users in the network. 
\begin{remark}
	In Theorem \ref{Theorem:1}, we analyze the SIR of the typical user located at the origin by using the Palm distribution of the user point process. Since the user point process is time invariant ergodic Poisson point process, the obtained formula corresponds to the spatial average of SIRs of all the users in the network \cite{baccelli2010stochastic,baccelli2010stochasticvol2}. In other words, it corresponds to the statistic of the SIRs of all users in a large ball, at any given time.  In addition, since the user point process is a time-invariant and ergodic Poisson point process, the coverage probability of the typical user coincides with the time average of the coverage probability of a specific user, obtained over a very long time \cite{baccelli2013elements}.
\end{remark}

\begin{figure}
	\centering
	\includegraphics[width=1\linewidth]{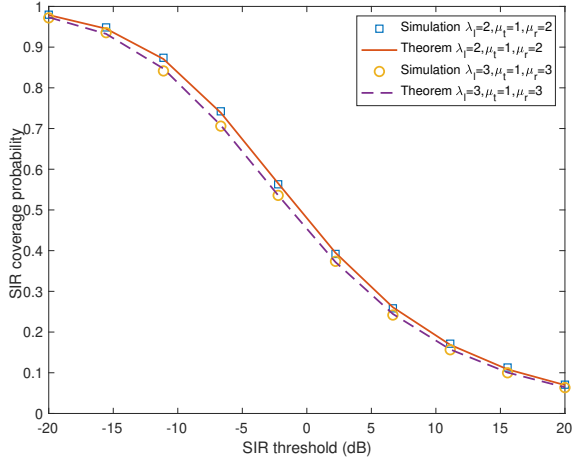}
	\caption{The derived formula matches the simulation results. We use $ \gamma =  1 $, $ \alpha=2.5  $ and $ \beta = 3.5 $.  The units of $ \lambda_l, \mu_s, $ and $\mu_r$ are per kilometer.}
	\label{fig:sir2122535}
\end{figure}
%
%\begin{figure}
%	\centering
%	\includegraphics[width=1\linewidth]{SIR_322254_323254_324254}
%	\caption{SIR coverage probability of the typical user. We consider $ \gamma =  1 $, $ \alpha=2.5  $ and $ \beta = 4 $.}
%	\label{fig:sir322254323254324254}
%\end{figure}

\begin{figure}
	\centering
	\includegraphics[width=1\linewidth]{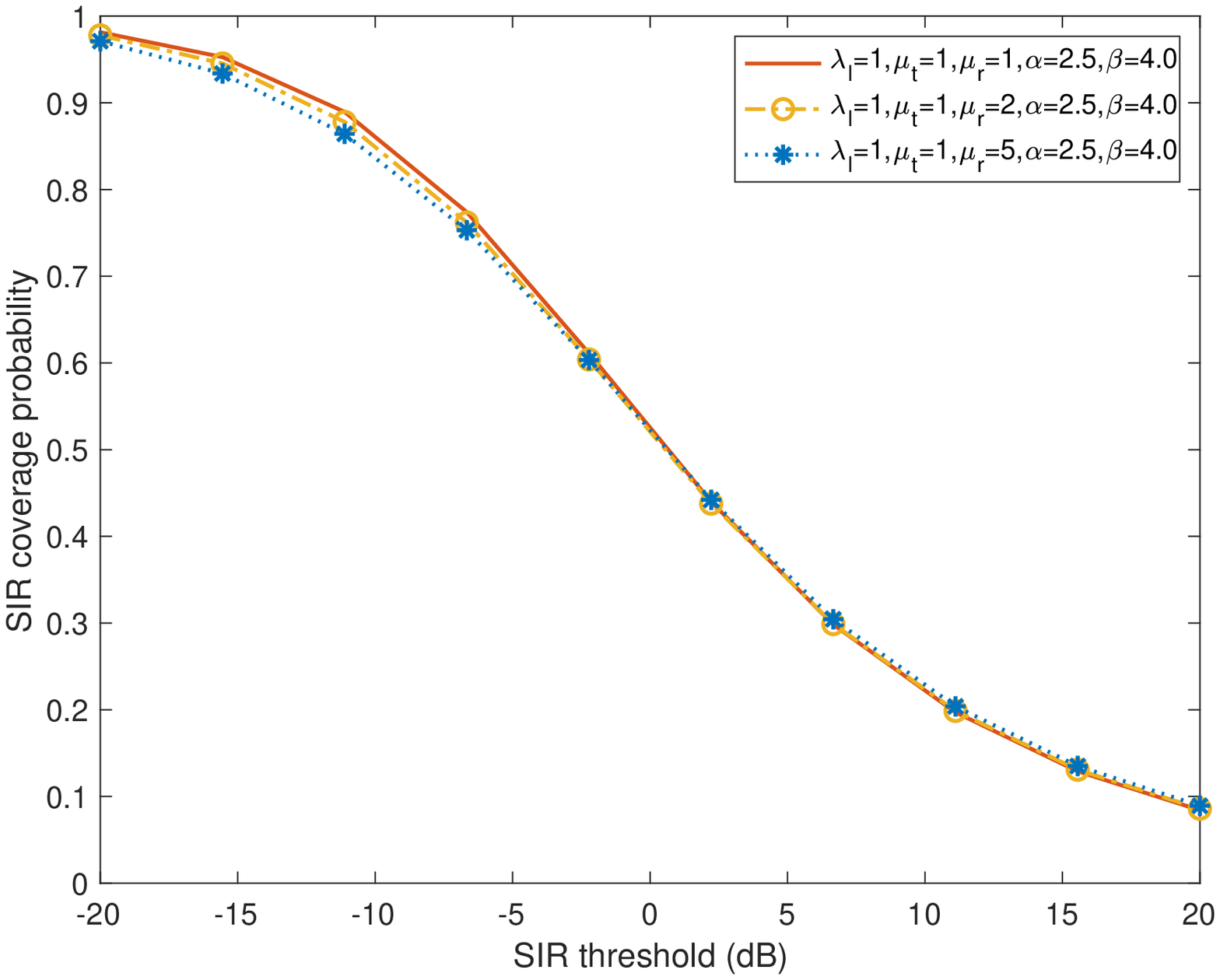}
	\caption{The illustration of the SIR coverage probability. Here, $ \lambda_l $ and $ \mu_s $ are fixed whereas $ \mu_r $ varies. }
	\label{fig:sir111254112254115254}
\end{figure}

\begin{figure}
	\centering
	\includegraphics[width=1\linewidth]{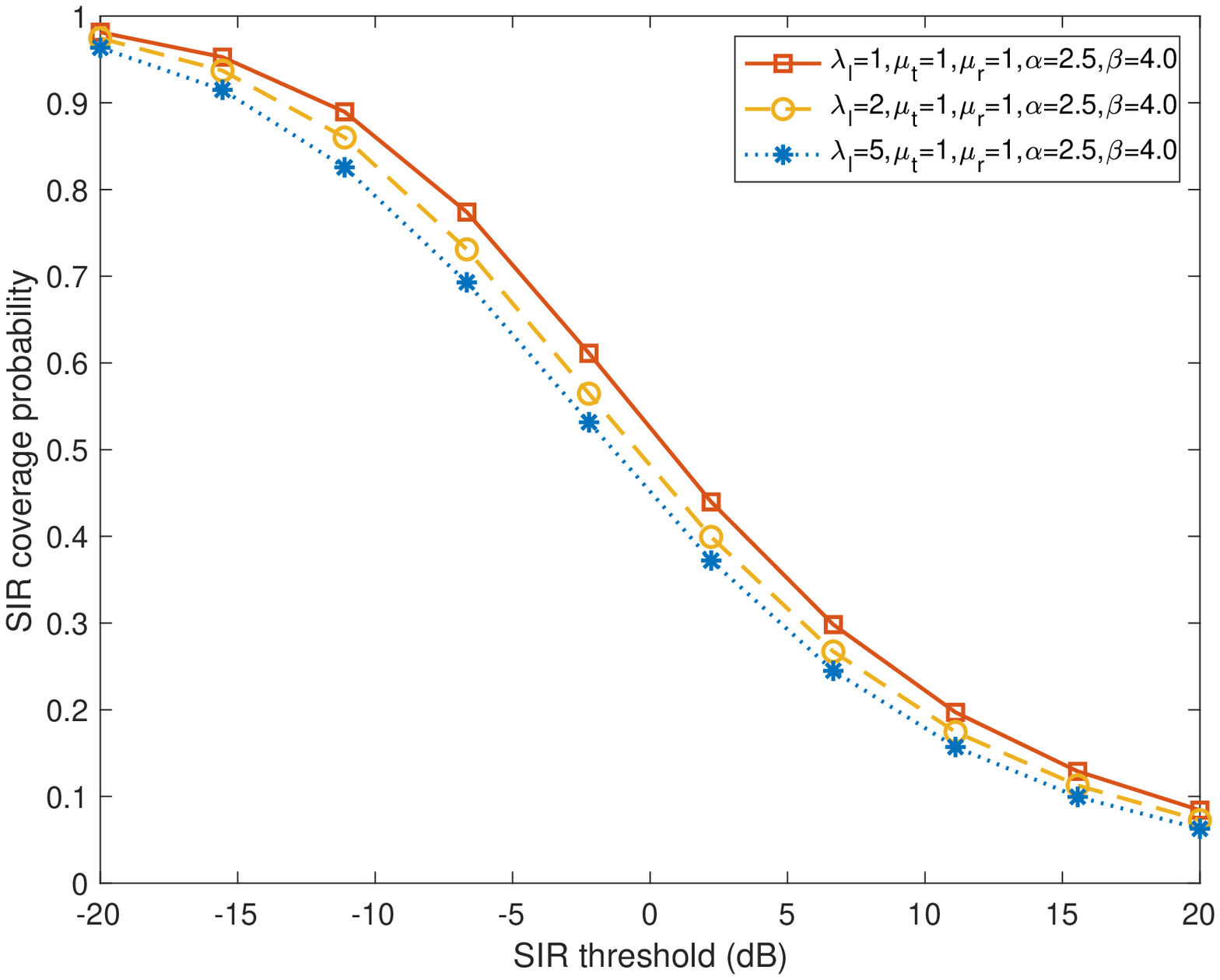}
	\caption{The illustration of the SIR coverage probability. Here,  $ \mu_s $ and $ \mu_r $ are fixed whereas $ \lambda_l $ varies. }
	\label{fig:sir1112541122541152542}
\end{figure}

\begin{figure}
	\centering
	\includegraphics[width=1\linewidth]{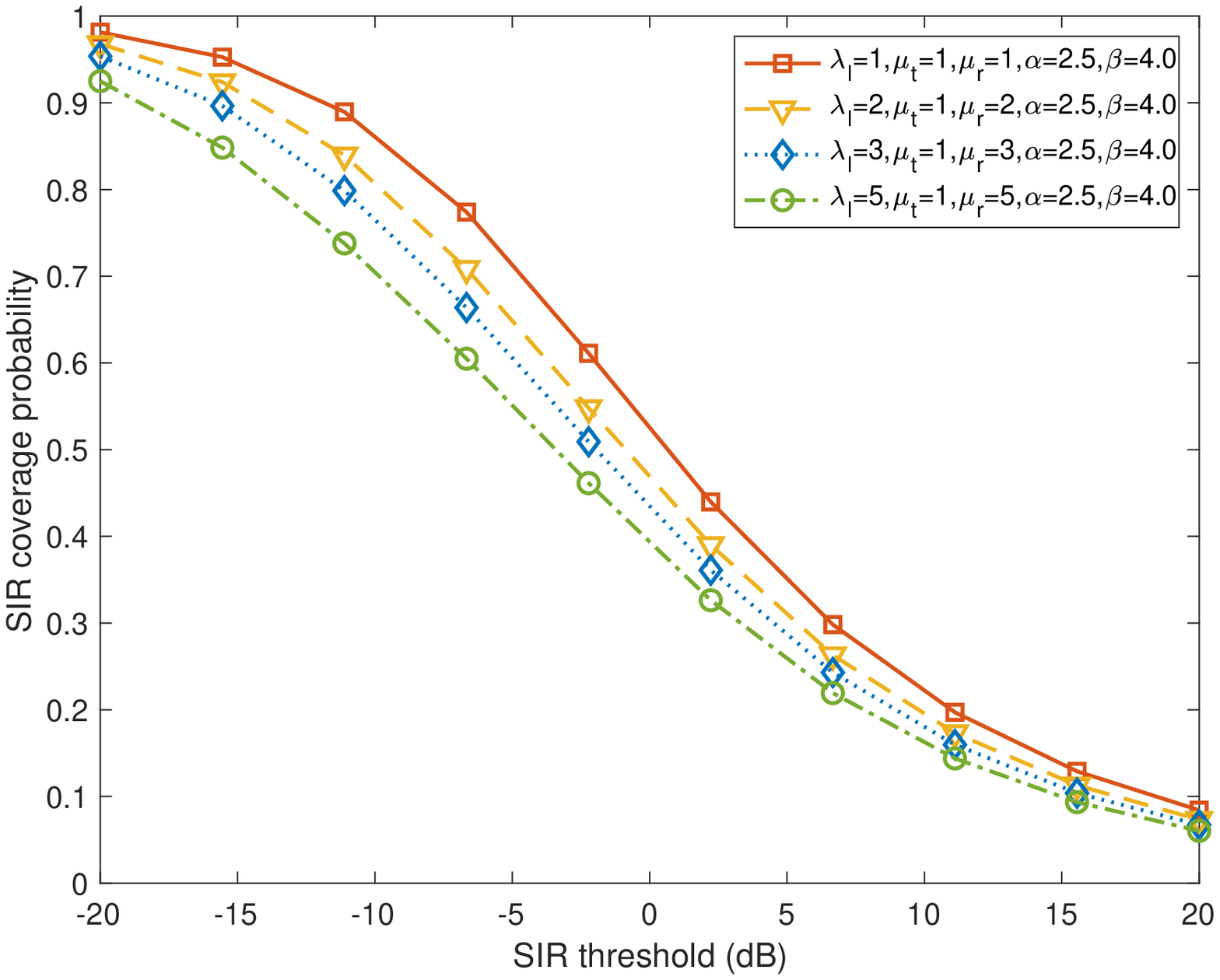}
	\caption{The SIR coverage probability. We use  $ \alpha\neq \beta $. }
	\label{fig:sir111254212254313254515254}
\end{figure}

Fig. \ref{fig:sir2122535} shows that the derived coverage probability of the typical user matches the numerical results obtained by Monte Carlo simulations, performed under various network parameters. In Figs. \ref{fig:sir111254112254115254} -- \ref{fig:sir111254212254313254515254}, we show only analytical results. Note that in the top figure of Fig. \ref{fig:sir111254112254115254}, the SIR curve slightly changes as the density of the relays varies. In the low SIR regime, the SIR curve slightly decreases as we increase the number of relays. This is because, in the low SIR regime, users are more likely to be associated with transmitters on lines that are different from the ones of users, and the received signal powers from the association transmitters are  moderately dominated by the interference from the other transmitters. On the other hand, in the high SIR regime, the SIR curve increases as the number of relays increases. This is because, in the high SIR regime, users are more likely to be associated with the transmitters on the lines that are the same as the ones of users, and therefore the received signal powers dominate the interference. Nevertheless, it is worthwhile to mention that increasing the relay density does not always increase the SIR curve in some range of parameters. Especially when the relay or RSU density is very high, transmitters and receivers may be very close to each other and thus the power-law path loss function of this paper should be replaced with a truncated version, e.g., $ L(d)=\min\{1,d^{-\alpha}  \text{ or } d^{-\beta}\} $, to account for near field effect. The analysis with a truncated power-law path loss model is left for future work.  In the right picture of Fig. \ref{fig:sir111254112254115254}, we increase the line density to show the change of the SIR curve. Although $ \mu_s $ and $ \mu_r $ are equal to one, the number of RSUs or relays $ \lambda_l\mu_s/\pi $ or $ \lambda_l\mu_r\pi $ respectively increases as $ \lambda_l $ increases. Therefore, the increment of the interference dominates the increment of the received signal power and this explains the decrement of the SIR curve as $ \lambda_l $ increases.

\begin{figure}
	\centering
	\includegraphics[width=1\linewidth]{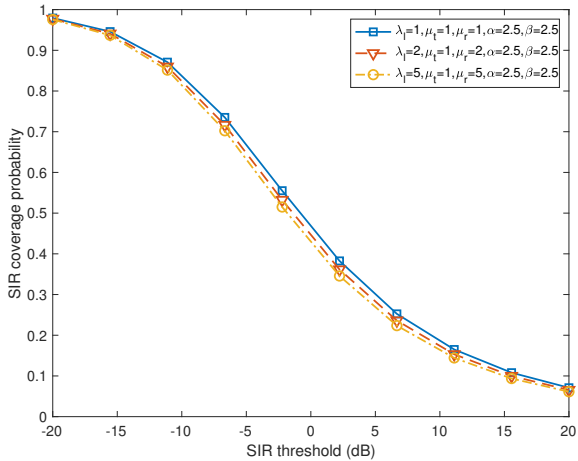}
	\caption{The SIR coverage probability. We use  $ \alpha = \beta $. }
	\label{fig:sir1112542122543132545152542}
\end{figure}

In Fig. \ref{fig:sir111254212254313254515254}, we increase the road density $ \lambda_l $ and the linear density of relays $ \mu_r $ at the same rate. In both of pictures, increasing the road density decreases the SIR curve.  It is important to mention that the SIR curve decrease in $ \alpha = \beta  $ is less significant than the SIR curve decrease in $ \alpha\neq \beta.$ By comparing the top figures of Fig. \ref{fig:sir111254112254115254} and  \ref{fig:sir111254212254313254515254}, we see that the SIR curve decrease much clearer in Fig. \ref{fig:sir111254212254313254515254} because, in general, the average number of transmitters per unit area is $ \lambda_l(\mu_s+\mu_v)/\pi $ and thus the top figure of Fig. \ref{fig:sir111254212254313254515254} has much more transmitters than the top figure of Fig. \ref{fig:sir111254112254115254}, on average. We can conclude that the interference caused by relays is significant for dense urban areas where roads are densely distributed. Nevertheless, by comparing $ \lambda_l=1,\mu_s=1,\mu_r=5 $ and $ \lambda_l=5,\mu_s=1,\mu_r=5 $, we see that the SIR curve decrease from $ \mu_r=1 $ to $ \mu_r=5 $ is about $ 15  $ -- $ 20  $\% when the SIR threshold is between $ -10 $ dB and $ 0  $ dB. When the SIR threshold is not within this range, the decrease is between $ 5 $ -- $ 10 $\%. From these observations, we conclude that despite some SIR decrease, relays are able to redistribute the users that are previously associated with RSUs. In the right picture of Fig. \ref{fig:sir111254212254313254515254}, users are more likely to be associated with relays as the relay density increases. Especially, in the low SIR regime,  users are associated with relays on different roads. Consequently, if the cross-road attenuation is not very significant, the received signal power from the association relays increases as we increase the number of relays and it compensates the interference from added relays to some extent. It is worthwhile to stress that such a behavior of the SIR curves exists as long as the density of RSUs or relays is not too high. For instance, a truncated path loss model should be used if transmitters and receivers are too close to each other.

\begin{corollary}\label{Cor:1}
	When $  p_s = p_r  $, or $ \gamma = 1 $, the coverage probability of the typical user is given by Eq. \eqref{eq:39} 	where 
	\begin{align*} 
		J_1(r)=2(\mu_s+\mu_r) e^{-2r(\mu_s+\mu_r)-2(\mu_s+\mu_r)\int_{r}^\infty \frac{\tau r^\alpha u^{-\alpha}}{1+\tau r^\alpha u^{-\alpha}}\diff u},
	\end{align*}
	\begin{align*}
		J_2(r,v)=& e^{-2(\mu_s+\mu_r)\sqrt{r^2-v^2}} \\
		&e^{-2(\mu_s+\mu_r) \int_{\sqrt{r^2-v^2}}^{\infty}\frac{\tau r^\alpha {(v^2+u^2)}^{-\frac{\beta}{2}}}{1+\tau r^\alpha {(v^2+u^2)}^{-\frac{\beta}{2}}}\diff u},
	\end{align*}
	\begin{align*}
		&J_3(r)=e^{-2(\mu_s+\mu_r) \int_{0}^{\infty}\frac{\tau r^\alpha {(v^2+u^2)}^{-{\beta}/{2}}}{1+\tau r^\alpha {(v^2+u^2)}^{-{\beta}/{2}}}\diff u},
	\end{align*}
	and 
	\begin{align*}
		&K_1(r)= e^{{-2(\mu_s+\mu_r) r}- 2(\mu_s+\mu_r) \int_r^\infty \frac{\tau r^\beta u^{-\alpha}}{1+\tau r^\beta u^{-\alpha}}\diff u},\\
		&K_2(r,v)= e^{-2(\mu_s+\mu_r)\sqrt{r^2-v^2}}\nnb\\
		&\hspace{18mm}e^{-2(\mu_s+\mu_r) \int_{\sqrt{r^2-v^2}}^{\infty}\frac{\tau r^\beta {(v^2+u^2)}^{-{\beta}/{2}}}{1+\tau r^\beta {(v^2+u^2)}^{-{\beta}/{2}}}\diff u},\\
		&K_3(r)=e^{-2(\mu_s+\mu_r) \int_{0}^{\infty}\frac{\tau r^\beta{(v^2+u^2)}^{-{\beta}/{2}}}{1+\tau r^\beta {(v^2+u^2)}^{-{\beta}/{2}}}\diff u},\\
		&K_4(r)= \int_0^{\pi/2}{4}\lambda_l (\mu_s+\mu_r)re^{-2(\mu_s+\mu_r)r\sin(\theta)}\nnb\\
		&\hspace{15mm}e^{-2(\mu_s+\mu_r)\int_{r\sin(\theta)}^{\infty}\frac{\tau r^\beta(r^2\cos^2(\theta)+v^2)^{-{\beta}/{2}}}{1+\tau r^\beta(r^2\cos^2(\theta)+v^2)^{-{\beta}/{2}}}\diff v}\diff \theta.
	\end{align*}
\end{corollary}
\begin{IEEEproof}
	Having $  \gamma = 1 $ in Theorem \ref{Theorem:1} completes proof. 
\end{IEEEproof}

\subsection{Coverage Probability of the Typical Relay}\label{S:33}
In practice, relays can serve users only when the relevant data are channeled through RSU-to-relay links. To evaluate the network user performance restricted by the RSU-to-relay links, this section evaluate the coverage probability of the typical relay. 

\begin{theorem}\label{Proposition:1}
	The coverage probability of the typical relay is given by Eq. \eqref{eqtheorem:2} where 
	\begin{align*} 
		&\bar{J}_1(r)=2\mu_s e^{-2r\mu_s-2\mu_s\int_{r}^\infty \frac{\tau r^\alpha u^{-\alpha}}{1+\tau r^\alpha u^{-\alpha}}\diff u},\nnb\\
		&\bar{J}_2(r,v)= e^{-2\mu_s\sqrt{r^2-v^2}-2\mu_s \int_{\sqrt{r^2-v^2}}^{\infty}\frac{\tau r^\alpha {(v^2+u^2)}^{-\frac{\beta}{2}}}{1+\tau r^\alpha {(v^2+u^2)}^{-\frac{\beta}{2}}}\diff u},\\
		&\bar{J}_3(r)=e^{-2\mu_s \int_{0}^{\infty}\frac{\tau r^\alpha {(v^2+u^2)}^{-{\beta}/{2}}}{1+\tau r^\alpha {(v^2+u^2)}^{-{\beta}/{2}}}\diff u},
	\end{align*}
	and   
	\begin{align*}
		&\bar{K}_1= e^{{-2 r\mu_s}- 2\mu_s \int_r^\infty \frac{\tau r^\beta u^{-\alpha}}{1+\tau r^\beta u^{-\alpha}}\diff u}, \nnb\\
		&\bar{K}_2(r,v)= e^{-2\mu_s\sqrt{r^2-v^2}-2\mu_s \int_{\sqrt{r^2-v^2}}^{\infty}\frac{\tau r^\beta {(v^2+u^2)}^{-{\beta}/{2}}}{1+\tau r^\beta {(v^2+u^2)}^{-{\beta}/{2}}}\diff u},\\
		&\bar{K}_3(r)=e^{-2\mu_s \int_{0}^{\infty}\frac{\tau r^\beta{(v^2+u^2)}^{-{\beta}/{2}}}{1+\tau r^\beta {(v^2+u^2)}^{-{\beta}/{2}}}\diff u},\\
		&\bar{K}_4(r)= \int_0^{\pi/2}{4}\lambda_l\mu_s r e^{-2\mu_sr\sin(\theta)} \nnb\\
		&\hspace{25mm}e^{-2\mu_s\int_{r\sin(\theta)}^{\infty}\frac{\tau r^\beta(r^2\cos^2(\theta)+v^2)^{-{\beta}/{2}}}{1+\tau r^\beta(r^2\cos^2(\theta)+v^2)^{-{\beta}/{2}}}\diff v}\diff \theta.
	\end{align*}
\end{theorem}

\begin{IEEEproof}
	We have the result by using techniques in the proof of Theorem \ref{Theorem:1}.  
\end{IEEEproof}
We combine Theorem \ref{prop:1}, Lemma \ref{Theorem:2}, Theorems \ref{Theorem:1}, and \ref{Proposition:1} to derive the user throughput.

\section{User Throughput}\label{S:34}
In the proposed network, users are associated with either RSUs or relays. 

Firstly, the normalized achievable rate of RSU-associated users is defined by the mean achievable rate of the typical RSU-associated users, divided by the mean number of users per RSU. The normalized achievable rate of the RSU-associated user is given by 
\begin{equation}
	\cT_{s}={\frac{W_1\bE[\log_2(1+\SIR_{s\to u})]}{\bE[\text{\# users per RSU}]}}\label{e_rate_R_to_u}. %_{\text{effective rate from RSU to user}}
\end{equation}
The normalized achievable rate is in a heuristic metric because it is given by the ratio of the achievable rate to the average number of users, not the exact number.  However, the exact distribution of the Cox-Voronoi cell is unknown; and thus using the exact number of users per RSU is infeasible. Here, we leverage the mass transport principle to obtain the mean number of users in the typical RSU cell (Theorem \ref{prop:1}) and use it to compute the normalized achievable rate of RSU-associated user. 

\par Secondly, the normalized achievable rate  of relay-associated users is dictated by both the coverage probabilities of the RSU-to-relay and relay-to-user links. Using the coverage probabilities of the both links, the normalized achievable rate of the relay-associated user is defined by 
% SIR-based achievable rates divided by the number of relays or by the number of users, respectively. Since the exact numbers of relays or users are random variables,  we use their averages in place. For instance, the average effective rate of RSU-to-relay link is given by the achievable rate of the typical RSU-to-relay link divided by the mean number of relays. As a result, the average effective rate of the relay-associated user is given by  
\begin{align}
	\cT_{r}=&\min\left\{{\frac{W_2\bE[\log_2(1+\SIR_{s\to r})]}{\bE[\text{\#  relay per  RSU}]\bE[\text{\#  user per  relay}]}}\right., \nnb\\ &\hspace{11mm}\left.{\frac{W_1\bE[\log_2(1+\SIR_{r\to u})]}{\bE[\text{\# user per relay}]}}\right\}\label{e_rate_r_to_u},
\end{align}
where  $ W_2 $ is the bandwidth for the RSU-to-relay links and $ W_1 $ is the bandwidth for RSU-to-user and relay-to-user links. 

Finally, $	\cT = \bP_U^0(A_s)\cT_{s}+\bP_U^0(A_r)\cT_{{r}} \label{e_rate_total}$
where $ \bP_U^0(A_s) $ and $ \bP_{ U }^0(A_r) $ are given by Theorem \ref{Theorem:2}. 

		\begin{remark}
	The instantaneous SIRs of RSU-to-relays links do not directly dictate the user throughput. However, these links indirectly affect the user performance by restricting the amount of data available at relays. Consequently, the throughput of relay-associated users will be determined by (i) the throughput of RSU-to-relay links, (ii) the throughput of relay-to-user links, (iii) the bandwidths $ W_1 $ and $ W_2, $ and (iv) the number of relays per RSU and the number of users per relay. 
\end{remark}

\begin{theorem}\label{Theorem:4}
	The user throughput is given by 
	\begin{align}
\cT=		&\bP_U^0( A_s)W_1\int_0^{\infty}\frac{\bP_U^0(\SIR_{s\to u}>2^{\xi}-1)}{\bar{u}_s} \diff \xi \nnb\\
		&+ \bP_U^0( A_r)\min\left\{\int_0^{\infty}\frac{W_2\bP_R^0(\SIR_{s\to r}>2^{\xi}-1)}{\bar{r}_s\bar{u}_r} \diff \xi ,\right. \nnb\\
		&\hspace{24mm}\left.\int_0^{\infty}\frac{W_1\bP_U^0(\SIR_{r\to u}>2^{\xi}-1)}{\bar{u}_r} \diff \xi \right\},\nnb
	\end{align}
where $ \bP_U^0( A_s) $ and $ \bP_R^0(\SIR_{s\to r}>2^{\xi}-1) $ are given by  Theorems \ref{Theorem:2} and \ref{Proposition:1}, respectively. Using the functions in Theorem \ref{Theorem:1}, the coverage probability of the RSU-associated typical user is given by Eq. \eqref{eq:RSU-}. Similarly, the coverage probability of the relay-associated typical user is given by Eq. \eqref{eq:66}. Using Theorem \ref{prop:1}, we have $\bar{u}_s= {\mu_u}\bP_U^0(A_s)/{\mu_s},$  $\bar{u}_r= {\mu_u}\bP_U^0(A_r)/{\mu_r}, $ and $
	\bar{r}_s={\mu_r}/{\mu_s}$ 
\end{theorem}

\begin{figure*}
		\begin{align}
		&\int_0^\infty \bar{J}_1(r)e^{-2\lambda_l\int_{0}^r 1-\bar{J}_{2}(r,v)\diff v }e^{-2\lambda_l\int_r^\infty 1-\bar{J}_{3}(r,v)\diff v }\diff r \nnb\\&\hspace{15mm}+ \int_{0}^\infty \bar{K}_{1}^{}(r)e^{-2\lambda_l\int_0^r 1-\bar{K}_2(r,v)\diff v-2\lambda_l\int_r^{\infty} 1-\bar{K}_3(r,v)\diff v}\bar{K}_{4}(r)\diff r.\label{eqtheorem:2}\\
%	\end{align}
%	\begin{align}
		& \frac{1}{\bP_U^0(\cA_s)} \left.\int_{0}^{\infty}G_1(r,a,b)e^{-2\lambda_l\int_{0}^r1-G_2(r,v,a,b)\diff v -2\lambda_l\int_{r}^{\infty}1-G_3(r,v,a,b)\diff v} \diff r \right|_{a=1,b=\frac{1}{\gamma}}\nnb\\
		&\hspace{15mm}+ \frac{1}{\bP_U^0(\cA_s)}\left.\int_{0}^\infty {H}_1(r,a,b)e^{-2\lambda_l\int_0^r 1-{H}_2(r,v,a,b)\diff v-2\lambda_l\int_r^{\infty} 1-{H}_3(r,v,a,b)\diff v}{H}_4(r,a,b)\diff r\right|_{a=1,b=\frac{1}{\gamma}, c=\mu_s}.\label{eq:RSU-}\\
%	\end{align}
%	\begin{align}
		&\frac{1}{\bP_U^0(\cA_r)}\left. \int_{0}^{\infty}{G}_1(r,a,b)e^{-2\lambda_l\int_{0}^r 1-{G}_2(r,v,a,b)\diff v }e^{-2\lambda_l\int_{r}^{\infty} 1-{G}_3(r,v,a,b)\diff v} \diff r\right|_{a=\gamma, b=1} \nnb\\
		&\hspace{15mm}+ \frac{1}{\bP_U^0(\cA_r)}\left.\int_{0}^\infty {H}_1(r,a,b)e^{-2\lambda_l\int_0^r 1-{H}_2(r,v,a,b)\diff v-2\lambda_l\int_r^{\infty} 1-{H}_3(r,v,a,b)\diff v}\bar{H}_4(r,a,b)\diff r\right|_{a=\gamma,b=1,c=\mu_r}.\label{eq:66}
	\end{align}
	\rule{\linewidth}{0.2mm}
\end{figure*}

\begin{IEEEproof}
	The coverage probabilities of the typical RSU-to-user link and of the typical relay-to-user link are obtained by leveraging Theorem \ref{Theorem:1}, respectively.
To obtain $ \bar{u}_s $,$ \bar{u}_r $, and $ \bar{r}_s ,$ we use Theorem \ref{prop:1}.  This completes the proof.
\end{IEEEproof}
\begin{figure}
	\centering
	\includegraphics[width=1\linewidth]{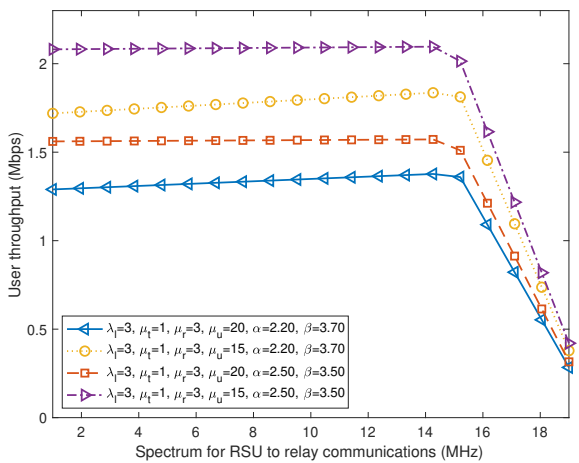}
	\caption{User throughput in Theorem \ref{Theorem:4}. }
	\label{fig:aergain}
\end{figure}
\begin{example}\label{ex:3}
	Suppose $ \gamma=1 $ and $ W_2$ is sufficiently large. Then, the user throughput is 
		\begin{align}
			\cT=&W_1\int_{0}^{\infty}\frac{\mu_s\bP_U^0(\SIR_{S\to U}>2^\xi -1)}{\mu_u}\diff \xi \nnb\\
			&\hspace{10mm}+ 	W_1\int_{0}^{\infty}\frac{\mu_r\bP_U^0(\SIR_{R\to U}>2^\xi -1)}{\mu_u}\diff \xi.		\label{43}
		\end{align}
	On the other hand, the user throughput without any relay  is
 \begin{equation}{\left(\frac{\mu_s}{\mu_u}\right)	W_1\int_0^\infty \bP_U^0(\SIR>2^\xi-1)\diff \xi.}\label{44}
 \end{equation}    
As a result, based on Eqs. \eqref{43} and \eqref{44}, the proposed network has a multiplicative gain $ \Gamma $ in the user throughput given by
\begin{align*}
	\Gamma =&\frac{\int_0^\infty \bP_U^0(\SIR_{S\to U}>2^\xi-1)\diff \xi}{\int_0^\infty \bP_U^0(\SIR>2^\xi-1)\diff \xi} \\ &+\left(\frac{\mu_r}{\mu_s}\right)\frac{\int_0^\infty \bP_U^0(\SIR_{R\to U}>2^\xi-1)\diff \xi}{\int_0^\infty \bP_U^0(\SIR>2^\xi-1)\diff \xi}. %\left(\frac{1}{\bP_U^0(A_s)}+\dfrac{\mu_r}{\mu_s\bP_U^0(A_r)}\right)
\end{align*}
\end{example}

\begin{remark}\label{ex:4}
	Theorem \ref{Theorem:4} shows the user throughput as a function of $ W_1, W_2, $ and the distributions of $\SIR_{S\to U}, \SIR_{R\to U} ,$ and $ \SIR_{S\to R}. $ It shows when $ W_2 $ is large, deploying relays will increase the user throughput of the network or the normalized achievable rate of user. In many cases where  $ W_1 + W_2 = W $, the derived user throughput formula is useful to the understanding tradeoff relationship between network parameters and addressing design problems exist in heterogeneous vehicular networks. For instance, to maximize the user throughput, one can find the optimal density of relays for a given $ W_2 $. Similarly, when the density of relays is given, one can use the user throughput formula to study the impact of $ W_2 $. 
\end{remark}

Fig. \ref{fig:aergain} shows the user throughput in Theorem \ref{Theorem:4} where we use $ W= 20 \text{ MHz}$, $\lambda_l=3/\text{km}$, $\mu_s=1/\text{km}$, $\mu_r=3/\text{km}$, $\mu_u=15/\text{km}$, $\alpha=2.5,$ and $ \beta=3.5. $ It shows that for the given network parameters, $ W_2=14 $ maximizes the user throughput of the proposed two-tier heterogeneous vehicular network. Note the maximum value of $ W_2 $ varies depending on the network variables such as $ \lambda_l,\mu_s,\mu_r,\alpha$ and $\beta. $ In practice, by exploiting Theorem \ref{Theorem:4} network operators can easily find the optimal solution for $ W_2 $ with a marginally little computation cost. 

%In Examples \ref{ex:3} and \ref{ex:4}, we study the user throughput  when $ W_2 $ is very high or the width spectrum resource for RSU-to-relay communications is very large. In practice, $ W_1+W_2=W $ and $ W $ is finite. Therefore, one should use Theorem \ref{Theorem:4} to allocate spectrum resources so as to maximize the user throughput of the two-tier heterogeneous vehicular network with RSUs and relays. 

%\begin{figure}
%	\centering
%	\includegraphics[width=0.4\linewidth]{egain}
%	\caption{Approximate user effective rate gain in the proposed network.}
%	\label{fig:egain}
%\end{figure}

\section{Conclusion and Future Work}
Using stochastic geometry, this paper proposes and analyzes a novel two-tier heterogeneous vehicular network architecture where RSUs and vehicular relays serve network users. By assuming such vehicular relays are operated by RSUs and users are associated with either RSUs or relays, we derive the association probability of the network users. We find that the association probability is a nonlinear function of the RSU and relay densities because RSUs, relays, and users are all on roads. Then, we derive the coverage probability of the typical user and then obtain the user throughput. In particular, the user throughput incorporates the fact that RSUs operate relays and that the throughput of relay-associated users is dictated by the SIRs of RSU-to-relay and relay-to-user links and the corresponding bandwidths for those links. The paper gives practical insights on designing heterogeneous vehicular networks with RSUs and vehicular relays. By presenting the formulas for SIR and user throughput as network parameters, one can easily identify the complex interactions occur at the network and use these formulas to enhance reliability or to increase throughput. 
\par The present paper starts a new line of studies on heterogeneous vehicular networks. It provides a tractable model and a tool to analyzing the network performance. The analysis of this paper can be developed further by considering new and more practical components; for instance, the clustering of vehicles on roads can be represented by an independent cluster point process on roads. The analysis of the proposed two-tier heterogeneous vehicular network is applicable to the analysis of multi-tier vehicular networks where there are various types of network elements exist such as RSUs, relays, and IoT devices.

\appendix[Proof of Theorem \ref{Theorem:1}]

\begin{IEEEproof}
	Under the Palm distribution of the user point process, $ \bP_{U}^0(\cdot) $, there exist a typical user at the origin and a line $ l(0,\theta_0) $ containing the typical user. Here, $ \theta_0 $ is a uniform random variable between $ 0 $ and $ \pi. $ By the law of total probability, the coverage probability is given by 
	\begin{align}
		&\bP_{ U }^0(\SIR >  \tau)\nnb\\&=\bP_{ U }^0(\SIR >  \tau,E)+ \bP_{ U }^0(\SIR >  \tau, {E}^c)\label{eq:sum},
	\end{align}
	where we can write $ E : \{l_\star = l(0,\theta_0) \} $ and $ E^{c}: \{l_\star \neq l(0,\theta_0)\}. $  The former is the event that the line $ l_\star $ containing the association transmitter  is $ l(0,\theta_0)  $, the line that contains the typical user. 
	
	\begin{align}
		&\bP_{U}^{0}(\SIR>\tau, E) \nnb\\
		&= \bP_{U}^{0}(\SIR>\tau, E,  A_s) + \bP_{U}^{0}(\SIR>\tau, E,  A_r ),
	\end{align}
	where $  A_s $ and $ A_r $ are the events that the typical user is associated with its closest RSU and with its closest relay, respectively. Then, with  $ I $ the interference seen by the typical user, we have 
	\begin{align}
		&\bP(\SIR>\tau, E, A_s)\nnb\\
		&=\bP_{ U }^0(p_sHL(\|X^\star\|) > \tau I,E,A_s)\nnb\\
		&=\bP_{ U }^0(p_sH > \tau I\|X_\star\|^{\alpha},E,A_s)\nnb\\
		&=\bE_{ \Phi }\left[\bP_{ U }^0\left(\left.H>\frac{\tau I \|X_\star\|^{\alpha}}{p_s}, E,A_s \right| \Phi \right)\right]\nnb\\
		&=\bE_{ \Phi }\left[\int_{r=0}^{r=\infty}  \bP_{ U }^0\left( \left.H>\frac{\tau r^\alpha I }{ p_s }\right| E, A_s,  \|X_S^\star\|=r,  \Phi  \right)\right. \nnb\\
		&\hspace{22mm}\left.\bP(\|X_S^\star\|\in[r,r+\diff r), E, A_s |\Phi)\right]\label{eq:8},
	\end{align}
	where we express the probability as the conditional expectation w.r.t. $ \Phi $. Then, we write it as a conditional expectation w.r.t. the nearest RSU. We have  $ \bP(\|X_S^\star\|= r\diff r , E, A_s|\Phi) $ the conditional probability density function of the distance from the origin to the closest RSU.   
	\par In a similar way, we have 
	\begin{align} 
		&\bP(\SIR>\tau, E, A_r)\nnb\\
		&=\bE_{ \Phi }\left[\int_{r=0}^{r=\infty}  \bP_{ U }^0\left(\left.H>\frac{\tau r^\alpha I }{ p_r }\right| E , A_r,  \|X_R^\star\|=r,  \Phi \right) \right.\nnb\\
		&\hspace{22mm}\left.\bP(\|X_R^\star\|\in[r,r+\diff r), E, A_r|\Phi)\right] \label{eq:9}.
	\end{align} 
	In Eq. \eqref{eq:9}, $ \bP(\|X_R^\star\|\in [r,r+\diff r), E, A_r |\Phi)$ is the conditional probability density function of the distance from the origin to the nearest relay. Furthermore, the integrands of Eqs. \eqref{eq:8} and \eqref{eq:9} are 
	\begin{align}
		&\bP_{ U }^0\left(\left.H>\frac{\tau r^\alpha  I }{p_s}\right| E,A_s, r, \Phi \right)  \nnb\\
		&= \bE_{ U }^0\left[e^{-sI}|E,A_s,r, \Phi \right]|_{s=\tau r^\alpha p_s^{-1}}\label{eq:10},\\
		&\bP_{ U }^0\left(\left.H>\frac{\tau r^\alpha  I }{p_r}\right| E,A_r ,r, \Phi \right) \nnb\\
		&=\bE_{ U }^0\left[e^{-sI}|E,A_r,r, \Phi \right]|_{s=\tau r^\alpha p_r^{-1}},\label{eq:10-2}
	\end{align}
	respectively. We obtain \[ \bE_{ U }^0\left[e^{-sI}|E,A_s,r, \Phi \right] = \bE_{ U }^0\left[e^{-sI}|E,A_r,r, \Phi \right] \] from the independence of the Poisson point processes. The conditional Laplace transform of interference is given by 
	\begin{align}
		&\bE_{ U }^0\left[e^{-sI}| E, A_s,r, \Phi \right]\\
		&=\prod_{T_k\in S_{0,\theta_0}+R_{0,\theta_0}} ^{|T_k|>r}\left(\frac{1}{1+sp{\|T_k\|}^{-\frac{\beta}{2}} }\right)\nnb\\
		&\hspace{13mm}\prod_{r_i,\theta_i\in\Phi\setminus 0,\theta_0}\left(\prod_{T_k\in S_{r_i,\theta_i}+R_{r_i,\theta_i}} ^{|T_k|>r}\frac{1}{1+sp{\|T_k\|}^{-\frac{\beta}{2}} }\right)\nnb
	\end{align}
	where we use the Laplace transform of the exponential random variable and the fact that conditionally on the line process and conditionally on the association distance $ r, $ all RSUs and relays are at distances greater than $ r. $ Then, we have 
	\begin{align}
		&\bE_{ U }^0\left[e^{-sI}| E, A_s,r, \Phi \right]\nnb\\
		&= e^{-2\mu_s \int_r^\infty \frac{s p_s u^{-\alpha}}{1+s p_s u^{-\alpha}}\diff u-2\mu_r \int_r^\infty \frac{s p_r u^{-\alpha}}{1+s p_r u^{-\alpha}}\diff u}\nnb\\
		&\hspace{3mm}\times\prod_{r_i\in\Phi}^{|r_i|<r} \left(e^{-2\mu_s \int_{\sqrt{r^2-r_i^2}}^\infty \frac{s p_s  (r_i^2+u^2)^{-\frac{\beta}{2}}}{1+s p_s  {(r_i^2+u^2)}^{-\frac{\beta}{2}}}\diff u} \right.\nnb\\
		&\hspace{20mm}\left.e^{-2\mu_r \int_{\sqrt{r^2-r_i^2}}^\infty \frac{s p_r  (r_i^2+u^2)^{-\frac{\beta}{2}}}{1+s p_r  {(r_i^2+u^2)}^{-\frac{\beta}{2}}}\diff u}\right)\nnb\\
		%&\times\prod_{r_i}^{|r_i|<r}\exp\left(\right)\nnb\\
		&\hspace{3mm}\times \prod_{r_i\in\Phi}^{|r_i|>r}\left(e^{-2\mu_s \int_{0}^\infty \frac{s p_s  (r_i^2+u^2)^{-\frac{\beta}{2}}}{1+s p_s  {(r_i^2+u^2)}^{-\frac{\beta}{2}}}\diff u}\right.\nnb\\
		&\hspace{20mm}\left.e^{-2\mu_r \int_{0}^\infty \frac{s p_r  (r_i^2+u^2)^{-\frac{\beta}{2}}}{1+s p_r  {(r_i^2+u^2)}^{-\frac{\beta}{2}}}\diff u}\right).\label{eq:11}
	\end{align}
	where we use the facts that RSU and relay point processes on different lines are conditionally independent and that the distances from the origin to any RSU points $ $$ \{T_k\}_{k\in\bZ}\in S_{r_i,\theta_i}  $ are given by $ \{\sqrt{r_i^2+G_k^2}\}_{k\in\bZ} ,$ where $ \{G_k\}_{k\in\bZ}$ is the RSU Poisson point process on the real axis $ S_{0,0} $. 
	
	On the other hand, the probability density function in Eq. \eqref{eq:8} is given by 
	\begin{align}
		&\bP(\|X_S^\star\|\in[r,r+\diff r),E, A_s |\Phi) \nnb\\
		&=\frac{\partial (1-\bP(S_{0,\theta_0}(B_0(r))=0))}{\partial r}\diff r\nnb\\ &\hspace{3mm}\times\bP(R_{0,\theta_0}(B(r))=\emptyset) \!\!\!\prod_{r_i,\theta_i\in\Phi}\!\!\!\bP(S_{r_i,\theta_i}+R_{r_i,\theta_i}(B(r))=\emptyset)\nnb\\
		&=2 \mu_s e^{-2r\mu_s}\diff r e^{-2r\mu_r}\prod_{r_i,\theta_i\in\Phi}^{|r_i|<r}e^{-2(\mu_s+\mu_r)\sqrt{r^2-r_i^2}},\nnb
	\end{align}
	where we use the facts that (i) $ X_\star\in S_{0,\theta_0} $ and (ii) there is no point of $ R_{0,\theta_0}+S_{r_i,\theta_i}+R_{r_i,\theta_i} $ within a disk of radius $ r $ centered at the origin. The probability density function in Eq. \eqref{eq:9} is 
	\begin{align}
		&\bP(\|X_R^\star\|\in(r,r+\diff r),E, A_r |\Phi) \nnb\\&=2 \mu_r e^{-2r\mu_r }\diff re^{-2r\mu_s} \prod_{r_i,\theta_i\in\Phi}^{|r_i|<r}e^{-2(\mu_s+\mu_r)\sqrt{r^2-r_i^2}}.\label{eq:12}
	\end{align}
	To obtain the first part of Eq. \eqref{eq:sum}, we combine Eqs. \eqref{eq:9} -- \eqref{eq:12}. 
	\par Let us now evaluate the second part of Eq. \eqref{eq:sum}. By the law of total probability, the expression $ \bP_{U}^{0}(\SIR>\tau, E^{c}) $ is 
	\begin{align}
		& \bP_{U}^{0}(\SIR>\tau,  E^{c},A_s)+\bP_{U}^{0}(\SIR>\tau, E^{c},A_r)\label{eq:16}
	\end{align}
	where $ A_s $ and $ A_r $ denote the events that the typical user is associated with the RSU or with the relay, respectively. 
	Let $ l_\star $ denote the line of the association RSU transmitter. Then, by conditioning on $ \Phi,$ on $ l_\star $, and then $ \|X_S^\star\| $, we can write the first part of Eq. \eqref{eq:16} as follows: 
	\begin{align}
		&\bP_{U}^{0}(\SIR>\tau, E^{c}, A_s)\nnb\\
		%&=\bP_{ U }^0(p_sHL(\|X^\star\|) > \tau I,E^c, A_s)\nnb\\
		&=\bP_{ U }^0(p_sH > \tau I\|X_\star\|^{\beta},E^c,A_s)\nnb\\
		&=\bE_{ \Phi }\left[\bP_{ U }^0\left(\left.H>\frac{\tau I \|X_\star\|^{\beta}}{p_s},E^c,A_s \right| \Phi \right)\right]\nnb\\
		&=\bE_{ \Phi , l_\star}\left[\bP_{ U }^0\left(\left.H>\frac{\tau I \|X_S^\star\|^{\beta}}{p_s},E^c,A_s \right| l_\star, \Phi \right)\right]\nnb\\
		&=\bE_{\Phi,l_\star,r}\left[\bP_{ U }^0\left(\left.H>\frac{\tau r^\beta I}{p_s}\right|  E^c, A_s , r, l_\star  \Phi \right)\right], \label{17}
	\end{align}	
	where we write $ \|X_S^\star\|=r. $

	In a similar way, the second part of Eq. \eqref{eq:16} is given by 
	\begin{align}
		&\bP_{U}^{0}(\SIR>\tau, E^{c}, A_r)\nnb\\
		&=\bE_{\Phi,l_\star,r}\left[\bP_{ U }^0\left(\left.H>\frac{\tau r^\beta I}{p_r}\right|  E^c, A_r , r, l_\star  \Phi \right)\right], \label{18}
	\end{align}
	where we write $ \|X_R^\star\|=r. $

	By using the fact that $ H $ is an exponential random variable, the conditional probability of Eq. \eqref{17} is given by expression \eqref{eq:mei} where the distances from the origin to the points of the Poisson point process on line $ l(r_i,\theta_i) $ are represented by  $ \|r_i\vec{e}_1+ T_j\vec{e}_2\| $ where $ \vec{e}_{i,1} $ is an unit $1$ orthogonal vector from the origin to the line $ l(r_i,\theta_i) $ and $ \vec{e}_{i,2} $ is an unit $1$ vector, orthogonal to the vector $ \vec{e}_{i,1} $. Here, $ S_{0,0} $ is the RSU point process on the $ x $-axis and $ R_{0,0} $ is the relay point process on the $ x $-axis. By using the probability generating functional of the Poisson point process, we have 
	\begin{align}
		\cL_{I}(s)&=e^{\left(-2\mu_s\int_{r}^{\infty}\frac{ p_s s v^{{-\alpha}}}{1+ p_s sv^{-\alpha}}\diff v-2\mu_r\int_{r}^{\infty}\frac{ p_r s v^{{-\alpha}}}{1+ p_r sv^{-\alpha}}\diff v\right)}\nnb\\
		&\hspace{4mm}\times \prod_{r_i,\theta_i\in \Phi+ \delta_{r_\star,\theta_\star} }^{|r_i|<r}e^{\left(-2\mu_s\int_{\sqrt{r^2-r_i^2}}^{\infty}\frac{ p_s s(r_i^2+v^2)^{-\frac{\beta}{2}}}{1+ p_s s(r_i^2+v^2)^{-\frac{\beta}{2}}}\diff v\right)}\nnb\\
		&\hspace{4mm}\times \prod_{r_i,\theta_i\in\Phi+\delta_{r_\star,\theta_\star}}^{|r_i|<r}e^{\left(-2\mu_r\int_{\sqrt{r^2-r_i^2}}^{\infty}\frac{ p_r s(r_i^2+v^2)^{-\frac{\beta}{2}}}{1+ p_r s(r_i^2+v^2)^{-\frac{\beta}{2}}}\diff v\right)}\nnb\\
		&\hspace{4mm}\times \prod_{r_i,\theta_i\in \Phi }^{|r_i|>r}e^{\left(-2\mu_s\int_{0}^{\infty}\frac{ p_s s(r_i^2+v^2)^{-\frac{\beta}{2}}}{1+ p_s s(r_i^2+v^2)^{-\frac{\beta}{2}}}\diff v\right)}\nnb\\
		&\hspace{4mm}\times \prod_{r_i,\theta_i\in\Phi}^{|r_i|>r}e^{\left(-2\mu_r\int_{0}^{\infty}\frac{ p_r s(r_i^2+v^2)^{-\frac{\beta}{2}}}{1+ p_r s(r_i^2+v^2)^{-\frac{\beta}{2}}}\diff v\right)},\label{eq:19}
	\end{align}
	where the above five terms of Eq. \eqref{eq:19} correspond to the Laplace transforms of the interference of (i) RSU plus relay on the line $ l(0,\theta_0), $ (ii) RSU on the lines closer than $ r $, (iii) relay on the lines closer than $ r, $ (iv) RSU on the lines further than $ r $, and (v) relay on the lines further than $ r, $ respectively.   
	\par 	To obtain the conditional probability density function of the distance from the origin to its closest RSU in Eq. \eqref{17}, we use the facts that (i) $ X_S^\star $ is the closest to the origin, (ii) $ X_S^\star\in S_{r_\star,\theta_\star}$, and (iii) all the other RSU or relay point processes have no point in the disk of radius $ r $. Therefore, using the void probability of the Poisson point process, the conditional probability density function of the distance from the origin to its closest RSU in Eq. \eqref{17} is 
	\begin{align}
		&\bP_U^0(\|X_S^\star\|\in[r,r+\diff r),E^c,A_s|l^\star,\Phi)\nnb\\
		&=\frac{\diff }{\diff r}{\left(1-\bP\left(S_{r_\star,\theta_\star}(B(r))=\emptyset\right)\right)}\diff r\bP(R_{r_\star,\theta_\star}(B(r))=\emptyset)\nnb\\
		&\hspace{15mm}\times \left(\prod_{r,\theta}^{ \Phi +{l_0}}\bP(R_{r,\theta}+S_{r,\theta}(B(r))=\emptyset)\right)\nnb\\
		&=\frac{2\mu_sre^{-2(\mu_s+\mu_r)\sqrt{r^2-r_\star^2}}}{\sqrt{r^2-r_\star^2}}\diff r\prod_{r_i,\theta_i}^{ \Phi +\delta_{0,\theta_0}}e^{-2(\mu_s+\mu_r) \sqrt{r^2-r_i^2}}.\label{eq:20}
	\end{align}
	\par In a similar way, the conditional probability density function of the distance from the origin to its closest relay in Eq. \eqref{18} is given by 
	\begin{align}
		&\bP_U^0(\|X_S^\star\|\in [r,r+\diff r), E^c,A_r|l^\star,\Phi)\nnb\\
		&=\frac{2\mu_rre^{-2(\mu_s+\mu_r)\sqrt{r^2-r_\star^2}}}{\sqrt{r^2-r_\star^2}}\diff r\prod_{r_i}^{ \Phi +\delta_{0,\theta_0}}e^{-2(\mu_s+\mu_r) \sqrt{r^2-r_i^2}}.\label{eq:21} 
	\end{align}
	Finally, we combine Eqs. \eqref{eq:19} and \eqref{eq:20} then integrate the result w.r.t. $ l_\star $ and then w.r.t. $ \Phi $. First, to integrate w.r.t. $ l_\star, $ we combine all the functions w.r.t. $ l_\star $ to get the expression \eqref{eq:233}. Then, we combine Eqs. \eqref{eq:19},  \eqref{eq:20},  and \eqref{eq:233}. 
	
	Similarly, to obtain the second part of Eq. \eqref{eq:16}, we combine Eq. \eqref{eq:19} and \eqref{eq:21} and evaluate the functions w.r.t. $ l_\star $ to obtain Eq. \eqref{eq:23}. Then, we combine the rest of Eq. \eqref{eq:19}, \eqref{eq:21} and \eqref{eq:23} to complete the proof. 
\end{IEEEproof}

\section*{Acknowledgment}
The work of Chang-Sik Choi was supported in part by
the NRF-2021R1F1A1059666.
The work of Francois Baccelli was supported by the ERC NEMO
grant 788851 to INRIA.

	\bibliographystyle{IEEEtran}
	\bibliography{ref}

\end{document}